**Increased hepatic PDGF-AA signaling mediates liver insulin resistance in obesity associated type 2 diabetes**

**Amar Abderrahmani**[1,2*], **Loïc Yengo**[1*], **Robert Caiazzo**[3*], **Mickaël Canouil**[1*,] **Stéphane Cauchi**[1], **Violeta Raverdy**[3], **Valérie Plaisance**[1], **Valérie Pawlowski**[1], **Stéphane Lobbens**[1], **Julie Maillet**[1], **Laure Rolland**[1], **Raphael Boutry**[1], **Gurvan Queniat**[1], **Maxime Kwapich**[1], **Mathie Tenenbaum**[1], **Julien Bricambert**[1], **Sophie Saussenthaler**[4], **Elodie Anthony** [5], **Pooja Jha**[6], **Julien Derop**[1], **Olivier Sand**[1], **Iandry Rabearivelo**[1], **Audrey Leloire**[1], **Marie Pigeyre**[3], **Martine Daujat-Chavanieu**[7], **Sabine Gerbal-Chaloin**[7], **Tasnim Dayeh**[8], **Guillaume Lassailly**[9], **Philippe Mathurin**[9], **Bart Staels**[10], **Johan Auwerx**[6], **Annette Schürmann**[4], **Catherine Postic**[5], **Clemens Schafmayer**[11], **Jochen Hampe**[12], **Amélie Bonnefond**[1,2], **François Pattou**[3*], **Philippe Froguel**[1,2*]

[1]Univ. Lille, CNRS, Institut Pasteur de Lille, UMR 8199 - EGID, F-59000 Lille, France; [2]Department of Medicine, Section of Genomics of Common disease, Imperial College London, UK; [3]Univ. Lille, Inserm, CHU Lille, U1190 - EGID, F-59000 Lille, France; [4]Department of Experimental Diabetology, German Institute of Human Nutrition Potsdam-Rehbrüecke, Nuthetal and German Center for Diabetes Research (DZD), München-Neuherberg, Germany. [5]Inserm, U1016, Institut Cochin, Paris, France CNRS UMR 8104, Paris, France Université Paris Descartes, Sorbonne Paris Cité, Paris, France. [6]Laboratory of Integrative and Systems Physiology, École Polytechnique Fédérale de Lausanne, 1015 Lausanne, Switzerland. [7]INSERM U1183, Univ. Montpellier, UMR 1183, Institute for Regenerative Medicine and Biotherapy, CHU Montpellier, France; [8]Department of clinical science; Skane University Hospital Malmö, Malmö, Sweden; [9]Univ. Lille, Inserm, CHU Lille, U995 - LIRIC - Lille Inflammation Research International Center, F-59000 Lille, France; [10]Univ. Lille, Inserm, CHU

Lille, Institut Pasteur de Lille, U1011- EGID, F-59000 Lille, France; [11] Department of Visceral and Thoracic Surgery, University Hospital Schleswig-Holstein, Kiel, Germany; [12]Medical Department 1, Technische Universität Dresden (TU Dresden), Dresden, Germany.

*These authors equally contributed to the study

**Corresponding authors**:

Philippe Froguel, MD, PhD, p.froguel@imperial.ac.uk, and Amar Abderrahmani, PhD, amar.abderrahmani@univ-lille2.fr

## Abstract (281 words)

Type 2 diabetes (T2D) is closely linked with non-alcoholic fatty liver disease (NAFLD) and hepatic insulin resistance, but the involved mechanisms are still elusive. Using DNA methylome and transcriptome analyses of livers from obese individuals, we found that both hypomethylation at a CpG site in *PDGFA* (encoding platelet derived growth factor alpha) and *PDGFA* overexpression are associated with increased T2D risk, hyperinsulinemia, increased insulin resistance and increased steatohepatitis risk. Both genetic risk score studies and human cell modeling pointed to a causative impact of high insulin levels on *PDGFA* CpG site hypomethylation, *PDGFA* overexpression, and increased PDGF-AA secretion from liver. We found that PDGF-AA secretion further stimulates its own expression through protein kinase C activity and contributes to insulin resistance through decreased expression of both insulin receptor substrate 1 and of insulin receptor. Importantly, hepatocyte insulin sensitivity can be restored by PDGF-AA blocking antibodies, PDGF receptor inhibitors and by metformin opening therapeutic avenues. **Conclusion:** Therefore, in the liver of obese patients with T2D, the increased PDGF-AA signaling contributes to insulin resistance, opening new therapeutic avenues against T2D and NAFLD.

**Abbreviations**

**ABOS**: Atlas Biologique de l'Obésité Sévère

**BMI:** Body Mass Index

**BMIQ**: Beta-MIxture Quantile normalization

**DMRs**: Differentially Methylated Regions

**FDR**: False Discovery Rate

**GRS**: Genetic Risk Score

**GWAS:** Genome-wide association studies

**HCC**: Hepatocarcinoma

**IHH**: immortalized human hepatocytes (IHH)

**INSR**: Insulin Receptor

**IRS1**: insulin receptor substrate 1 (IRS1)

**NAFLD:** Non-Alcoholic Fatty Liver Disease

**NASH:** Non-Alcoholic Steato-Hepatitis

**PCA**: Principal Component Analysis

**PDGFA:** Platelet Derived Growth Factor A

**PMA**: Phorbol 12-myristate 13-actetate

**PKC**: Protein Kinase C

**SNPs**: Single Nucleotide Polymorphisms

**STKs**: Serine/threonine protein kinases

In type 2 diabetes (T2D), hepatic insulin resistance is strongly associated with non-alcoholic fatty liver disease (NAFLD), a condition that hits >70% of diabetic patients (1). However, the molecular mechanisms involved are still incompletely understood. Genome-wide association studies (GWAS) and related metabolic traits have identified many loci associated with the risk of T2D (2). However, these loci only explain 15% of T2D inheritance, and GWAS have opened limited insights into the pathophysiology of T2D and its related complications including NAFLD (2). Epigenetic analyses of liver have enabled the identification of novel metabolic players and associated mechanisms involved in hepatic insulin resistance and NAFLD (3). Further, several DNA methylome wide association studies have identified candidate genes possibly involved in metabolic dysfunction of the adipose tissue, skeletal muscle and liver in obesity and T2D (4–8).

In this study, we hypothesized that DNA methylome of livers from patients with T2D, when compared to livers of individuals with normal plasma glucose levels, can unveil key players of hepatic insulin resistance and NAFLD in response to a diabetogenic environment. Our DNA methylome and transcriptome wide association analyses for T2D in liver samples from obese subjects identified a reduced methylation of a CpG site within the platelet-derived growth factor A gene (*PDGFA*), which was consistent with an increase in *PDGFA* mRNA. *PDGFA* encodes a protein forming a PDGF-AA homodimer, which is known as a liver fibrosis factor when overexpressed in the liver (9). We found that the rise of liver *PDGFA* is not only associated with increased T2D risk, but also with increased steatohepatitis (NASH) risk, increased fasting insulin levels and increased insulin resistance. The increased *PDGFA* expression and PDGF-AA protein levels were reproduced in human hepatocytes made insulin-resistant by long-term insulin incubation, and were present in the liver of insulin-resistant rodents. Furthermore, we have demonstrated the role of elevated PDGF-AA in perpetuating hepatocyte insulin resistance via a vicious autocrine negative feedback loop.

**Experimental procedure**

**Discovery study.** Liver biopsies were collected from 192 subjects from the French obesity surgery. Subjects included in the discovery study were participants of the ABOS ("*Atlas Biologique de l'Obésité Sévère*") cohort (ClinicalGov NCT01129297) including 750 morbidly obese subjects whose several tissues were collected during bariatric surgery (10). All subjects were unrelated, women, above 35 years of age, of European origin verified by Principal Component Analysis (PCA) using SNPs on the Metabochip array, non-smoker, non-drinker, without any history of hepatitis, and without indications of liver damage in serological analysis (normal ranges of aspartate aminotransferase, alanine aminotransferase and gamma-glutamyl transpeptidase). Overall, 96 T2D cases and 96 normoglycemic participants were selected. Normoglycemia and T2D were defined using the World Health Organization/International Diabetes Federation 2006 criteria (Normoglycemia: fasting plasma glucose < 6.1 mmol/l or 2-h plasma glucose < 7.8 mmol/l; T2D: fasting plasma glucose ≥ 7 mmol/l or 2-h plasma glucose ≥ 11 mmol/l). Each participant of the ABOS cohort signed an informed consent. For calculation of intermediate metabolic traits (HOMA2-IR and HOMA2-B indexes), see the **supplemental information**. All procedures were approved by local ethics committees. The main clinical characteristics were presented in **Table S1.**

**Replication study.** The replication study was based on *in silico* data of liver samples analyzed by the Infinium HumanMethylation450 BeadChip, as previously reported (11). Clinical characteristics were reported in **Table S2**. Liver samples were obtained percutaneously from subjects undergoing liver biopsy for suspected nonalcoholic fatty liver disease or intraoperatively for assessment of liver histology. Normal control samples were recruited from samples obtained for exclusion of liver malignancy during major oncological surgery. None of the normal control subjects underwent preoperative chemotherapy, and liver histology demonstrated absence of both cirrhosis and malignancy. Consenting subjects underwent a routine liver biopsy during bariatric surgery for assessment of liver affection. Biopsies were immediately frozen in liquid nitrogen, ensuring an *ex vivo* time of less than 40 seconds in all cases. A percutaneous follow-up biopsy was obtained in consenting bariatric patients five to nine months after surgery. Patients with evidence of viral hepatitis, hemochromatosis, or alcohol consumption greater than 20 g/day for women and 30 g/day for men were excluded. All patients provided written, informed consent. The study protocol was approved by the institutional review board (''Ethikkommission der Medizinischen Fakultät der Universität Kiel,'' D425/07, A111/99) before the beginning of the study.

**Epigenome-wide DNA methylation profiling**. The epigenome-wide analysis of DNA methylation was performed using the Infinium HumanMethylation450 BeadChip (Illumina, Inc., San Diego, CA, USA) which interrogates 482,421 CpG sites and 3,091 non-CpG sites covering 21,231 RefSeq genes (12). We used 500ng of DNA from liver tissue for bisulfate conversion using the EZ DNA Methylation kit D5001 (Zymo Research, Orange, CA, USA) according to the manufacturer's instructions. For the details of procedure, see the **supplemental information.**

**SNP genotyping, ethnic characterization and genetic risk score.** SNP genotyping was performed with Metabochip DNA arrays (custom iSelect-Illumina genotyping arrays) using the Illumina HiScan technology and GenomeStudio software (Illumina, San Diego, CA, USA) (13). We selected SNPs with a call rate ≥ 95 % and with no departures from Hardy–Weinberg equilibrium ($p>10^{-4}$). A Principal Component Analysis (PCA) was performed in a combined dataset involving the 192 patients plus 272 subjects from the publicly available HapMap project database. For details of analysis, **see the supplemental information.**

**Statistical Analyses**. Statistical analysis and quality control were performed with R software version 3.1.1. Raw data (IDAT file format) from Infinium HumanMethylation450 BeadChips were imported into R using the *minfi* package (version 1.12.0 on Bioconductor) (14), then we applied the preprocessing method from GenomeStudio software (Illumina) using the reverse engineered function provided in the *minfi* package. Samples were excluded when less than 75 % of the markers had detection *p*-values below $10^{-16}$. Markers were ruled out when less than 95 % of the samples had detection *p*-values below $10^{-16}$. According to this strategy, no sample was excluded and 70,314 markers (over 485,512) were excluded. For correction for Infinium HumanMethylation450 BeadChip design which includes two probe types (Type I and Type II), a Beta-MIxture Quantile normalization (BMIQ) was performed (15). Moreover, we checked for outliers using Principal Component Analysis (PCA) (*flashpcaR* package, version 1.6-2 on CRAN). At this stage, 416,693 markers and 192 samples were kept for further analysis. To test the association between methylation level and diabetic status, we applied a linear regression adjusted for steatosis (in percent), presence of NASH and fibrosis. Results were corrected for multiple testing using a Bonferroni correction ($p<10^{-7}$). The association between DNA methylation and metabolic traits was analyzed using a linear regression model, including normoglycemic samples adjusted for age and BMI. Quality control was performed on the HumanHT-12 v4.0 Whole-Genome DASL HT Assay (Illumina) data, according to the following criterion: probes were kept for further analysis when the detection *p*-values provided by GenomeStudio software version 3.0 (Illumina) were below five percent for all samples. A PCA was performed to identify samples with extreme transcriptomic profiles. After the quality control just described, 18,412 probes matching 13,664 genes and 187 samples were kept and analyzed for differential expression between T2D cases and controls, using linear regression. To account for multiple testing, we used five percent as a threshold for false discovery rate (FDR). Methylation and expression data were tested for correlation.

We selected a subgroup of 24 samples among the 192 initial samples, including 12 normoglycemic and 12 T2D cases, to analyze DNA methylation in blood samples from the same donors. The 24 samples were selected based on their expression and methylation profiles using PCA to reduce the heterogeneity.

**Cell Culture.** Immortalized Human Hepatocytes (IHH) were cultured in Williams E medium (Invitrogen), containing 11 mM glucose and supplemented with 10 % fetal calf serum (FCS; Eurobio), 100 U/ml penicillin, 100 µg/ml streptomycin, 20 mU/ml insulin (Sigma-Aldrich) and 50 nM dexamethasone (Sigma-Aldrich) (16). For insulin pre-treatment, $10^6$ cells were cultured in 6-well plates in a Dulbecco's Modified Eagle Medium (DMEM; Invitrogen) with or without 100 nM human insulin (Novo Nordisk) supplemented with 5 mM Glucose, 2 % FCS, 100 U/ml penicillin, 100 µg/ml streptomycin for 24 hours. For monitoring insulin signaling, medium was removed and replaced by FCS- and phenol red-free DMEM medium with or without 200 nM human insulin for one hour. Human hepatocytes were isolated from liver lobectomies resected for medical reasons as described (17) in agreement with the ethics procedures and adequate authorization.

**RNA sequencing, Microarray mRNA expression analysis, qRT-PCR, Western Blotting, Chemicals, ELISA, Glycogen measurement, Global Serine/Threonine kinases activity, DNA/RNA preparation, Oil-Red staining, cell proliferation, apoptosis, intermediate metabolic traits.** See the supplemental experimental procedure in the supplemental information

**Results**

**Liver epigenetic modification in T2D**

The liver DNA methylome was assessed in 96 age and body mass index (BMI)-matched obese women with T2D and 96 obese women with normal glucose levels (**Table S1**). While we initially identified 381 differentially methylated regions (DMRs) in the liver from obese patients with T2D compared to normal glucose obese patients (**Fig. S1**), we only observed one genome-wide significant DMR (cg14496282 within *PDGFA*) in obese patients with T2D after adjusting for liver steatosis and NASH, in an attempt to control for liver cellularity and impairment severity. The methylation at cg14496282 was associated with decreased T2D risk ($\beta = -15.6$ %; $p = 2.5 \times 10^{-8}$; **Fig. 1a and 1b**). The average DNA methylation at the cg14496282 was 41.3 % in patients with T2D and 60.3 % in controls, which corresponds to a 1.46-fold decrease in the methylation level of the CpG site. We checked the methylation, at this CpG site, for possible confounding effects due to differences in cell composition (18), and we still observed consistent effects for T2D risk ($\beta = -14.9$ %; $p = 6.9 \times 10^{-7}$). We replicated this association in the liver from 12 German cases with T2D and 53 German control subjects (11), and found that T2D risk was associated with decreased methylation level at cg14496282 site ($\beta = -14.0$ %; $p = 0.01$) (**Table S2**). These data were also supported by a recent study showing a decrease in *PDGFA* methylation in the liver from obese men with T2D compared to non-obese controls (4).

We next investigated whether the T2D-associated *PDGFA* cg14496282 hypomethylation was specific to the liver. We assessed the blood DNA methylome from 12 obese cases with T2D and 12 obese normal glucose controls presenting with extreme liver methylation levels at cg14496282. We found a significant correlation between methylation levels in blood and liver ($r = 0.66$; $p = 6.61 \times 10^{-4}$), and a slightly reduced methylation at the cg14496282 site ($\beta = -1.4$ %; $p = 0.01$) in blood of subjects with T2D when compared to controls. We also compared DNA methylation at cg14496282 in 43 liver and skeletal muscle samples from 192 participants who were randomly selected, but we did not find any significant correlation ($p > 0.05$).

In the 192 obese liver samples, we next investigated *cis*-located genes (within 500 kb around cg14496282) that were differentially expressed between T2D cases and controls, and which mRNA expression correlated with DNA methylation at *PDGFA* cg14496282 site. Using a false discovery rate threshold of 5 % for differential expression analysis and methylation-expression correlation analysis, we identified that the methylation at cg14496282 is negatively associated with the expression of *PDGFA* in T2D cases and normal glucose controls ($p < 0.007$; **Table 1**).

**Reduced liver *PDGFA* expression is associated with lower hepatic fibrosis risk**

In subjects with T2D and in normoglycemic controls, we found that *PDGFA* cg14496282 methylation was significantly associated with decreased NASH risk ($p < 0.05$; **Table 1**), while *PDGFA* expression in the liver was associated with increased NASH risk ($p < 0.01$; **Table 1**). Furthermore, in patients with T2D, *PDGFA* cg14496282 methylation was significantly associated with decreased hepatic fibrosis, decreased alanine aminotransferase levels and decreased aspartate aminotransferase levels ($p < 0.05$; **Table 1**), while *PDGFA* expression in the liver was associated with increased hepatic fibrosis and increased liver enzyme levels ($p < 0.01$; **Table 1**). These results were in line with previous studies that showed that *PDGFA* cg14496282 hypomethylation is associated with increased *PDGFA* liver expression in advanced *versus* mild human NAFLD (19,20). *PDGFA* encodes a dimer disulfide-linked polypeptide (PDGF-AA) that plays a crucial role in organogenesis and cirrhotic liver regeneration (21,22). Overexpression of PDGF-AA in mice liver causes spontaneous liver fibrosis (9). Moreover, activation of PDGF receptor signaling stimulates hepatic stellate cells and thereby, promotes liver fibrosis (23–25).

**Increased liver *PDGFA* expression is associated with hyperinsulinemia and insulin resistance**

In obese subjects with normal glucose levels, we next found that *PDGFA* cg14496282 methylation is significantly associated with decreased fasting serum insulin levels and decreased insulin resistance as modeled by the homeostasis model assessment index HOMA2-IR ($\beta = -1.45 \times 10^{-3}$, $p = 2.32 \times 10^{-3}$; and $\beta = -0.10$, $p = 4.93 \times 10^{-3}$, respectively; **Table 1**). In contrast, *PDGFA* liver expression was significantly associated with increased fasting serum insulin levels and increased insulin resistance ($\beta = 6.83 \times 10^{-3}$, $p = 9.49 \times 10^{-3}$; and $\beta = 0.53$, $p = 7.47 \times 10^{-3}$, respectively; **Table 1**).

Subsequently, we calculated a genetic risk score (GRS) as the sum of alleles increasing fasting insulin levels over 19 GWAS-identified single nucleotide polymorphisms (SNPs) (26), and found that this GRS is associated with decreased DNA methylation at cg14496282 ($\beta = -1.05$ % per allele; $p = 4 \times 10^{-3}$; **Table S3**). This association remained significant when we analyzed T2D cases and controls separately (and then meta-analyzed) or when we adjusted for BMI, high-density lipoprotein (HDL) cholesterol or triglycerides; these traits having a genetic overlap with fasting insulin (26). These results strongly suggested that hyperinsulinemia (and associated insulin resistance) contributes to decreased DNA methylation of *PDGFA* cg14496282 and consequently to the increase in the *PDGFA* expression. In contrast, the GRS including 24 SNPs associated with fasting glucose, the GRS including 65 SNPs associated with T2D and the GRS including 97 SNPs associated with BMI were not associated with cg14496282 methylation (**Table S3**), which excluded that hyperglycemia and obesity *per se* are directly modulating *PDGFA* methylation and expression.

The *in vivo* association between liver *PDGFA* overexpression and insulin resistance was supported by data obtained in different mice models of insulin resistance associated with obesity. In the liver from C57BL/6J (B6) mice that are susceptible to diet-induced obesity (27), we found that *Pdgfa* expression is increased by 46 %, as compared with control mice (*i.e.* that do not respond to a high-fat diet) (**Fig. 1c**). Similarly, we found that liver *Pdgfa* expression is increased in insulin resistant BXD mice fed a high-fat diet for 21 weeks when compared to control mice (**Fig. 1d**). However, the cg14496282 CpG site is not conserved in mice (28), suggesting that different epigenetic mechanisms rely on the rise of *Pdgfa/PDGFA* in insulin-resistant hepatocytes in mice and humans.

**Increased *PDGFA* expression and secretion from insulin-resistant human hepatocytes**

The association of increased liver *PDGFA* expression with insulin resistance in obese subjects suggests that PDGFA overexpression plays a role in liver insulin resistance, and therefore in T2D development. Chronic hyperinsulinemia indeed induces liver insulin resistance (29) and PDGFA has an autocrine function on hepatocytes (20). In this context, the exposure of mouse embryo cells to PDGF-AA inhibits insulin signaling (30). Therefore, we hypothesized that the obesity-associated increased PDGFA expression contributes to mediate the deleterious effects of chronic hyperinsulinemia on hepatic insulin resistance. To assess this hypothesis, we established an *in vitro* model of insulin-resistant human hepatocytes caused by hyperinsulinemia. To do so, we used the immortalized human hepatocytes (IHH). IHH cells are indeed equipped will the functional machinery for glucose metabolism (15) and they secrete PDGF-AA homodimer at comparable levels with primary human hepatocytes (**Fig. 2a**). Exposure of IHH cells to insulin for 16 h or 24 h hampered insulin-induced phosphorylation of AKT serine/threonine kinase at residue serine 473 (**Fig. 2b**). In line with AKT activation pivotal role in glycogen synthesis (31), we found reduced insulin-induced glycogen production in IHH cells exposed to insulin for 16 h and 24 h (**Fig. 2c**). The defective insulin signaling by insulin was accompanied by a rise in *PDGFA* mRNA, and in abundance and secretion of the encoded protein PDGF-AA homodimer from IHH cells (**Fig. 2d-f**). In addition, the insulin-induced increase in *PDGFA* expression was associated with the reduced cg14496282 CpG methylation level in IHH cells (**Fig. 2g**), suggesting that insulin resistance induced by hyperinsulinemia accounts for the rise of *PDGFA* in hepatocytes of obese individuals with diabetes. As these results may be cell line dependent, we measured AKT phosphorylation in liver hepatocellular HepG2 cells exposed to insulin for 24 h and retrieved similar results (**Fig. 2h**). In HepG2 cells, the expression of *PDGFA* was also significantly increased after long-term insulin incubation (**Fig. 2i**). The increase in *PDGFA* mRNA seemed specific to insulin exposure as neither cell proliferation (**Fig. S2**) nor the intracellular neutral lipid levels was modified by insulin treatment (**Fig. S3**). On the other hand, palmitate exposure did not change *PDGFA* expression in IHH cells (**Fig. S3**). Our data suggest that it is chronic hyperinsulinemia and not the excess of fatty acid influx that induces PDGF-AA secretion in the liver of obese individuals with T2D.

**PDGF-AA contributes to insulin resistance induced by insulin**

We then hypothesized that PDGF-AA directly causes insulin resistance in human hepatocytes. In line with this hypothesis, we found that the culture of IHH cells with a human PDGF-AA recombinant inhibits insulin-induced AKT activation (**Fig. 3a**). In contrast, the incubation of IHH cells with anti-PDGF-AA blocking antibodies reversed the deleterious effect of insulin long term exposure on AKT phosphorylation (**Fig. 3b**). We then investigated the mechanism whereby PDGF-AA inhibits AKT activation. Human hepatocytes express PDGF receptors (PDGFR) including PDGFRα and PDGFRβ that both bind PDGF-AA (22). We tested the role of PDGFR signaling using the PDGFR inhibitor Ki11502 (32). Pre-treatment of IHH cells with Ki11502 efficiently antagonized the negative effect of insulin on AKT phosphorylation mimicking results obtained with PDGF-AA blocking antibodies (**Fig. 3c**).

PDGFR blockade by Ki11502 increased the ability of insulin to stimulate glycogen synthesis (**Fig. 3d**).

To further dissect the signaling pathways by which both insulin and PDGF-AA impair AKT and insulin action, we performed a global measurement of serine/threonine protein kinases (STKs) using STK PamGene arrays consisting of 140 immobilized serine/threonine-containing peptides that are targets of most known kinases (33). We looked for differential STK activity between control and IHH cells cultured with insulin for 24 h. Peptides whose phosphorylation varied significantly between the two conditions were indicative of differential specific STK activities. This unbiased kinase analyses underscored significant differences in protein kinases C (PKCθ and PKCε) activities (**Fig. 4a**). The activation of these two PKCs hampers insulin signaling in response to chronic hyperlipidemia (34–36). These two kinases are also known to phosphorylate the insulin receptor substrate 1 (IRS1) and the insulin receptor (INSR) on serine residues, that impairs the association of INSR with IRS proteins, leading to the blockade of AKT activation and of the downstream signaling pathways (35, 36). Therefore, we treated IHH cells with phorbol 12-myristate 13-actetate (PMA), a potent activator of PKCs, and retrieved AKT inhibition (**Fig. S4**). PKCθ and PKCε kinase activities are linked to their phosphorylation at Serine 676 and Serine 729, respectively (37,38). In IHH cells cultured with insulin for 16 h or 24 h, we found a striking phosphorylation of the two PKCs, which coincided with the decreased AKT phosphorylation (**Fig. 4b**). The effect of insulin on the phosphorylation of the two kinases is likely to rely on PDGF-AA, as the PKCθ and PKCε were directly activated by PDGF-AA (**Fig. 4c**). Activation of PKCε decreases INSR abundance (34). In line with this result, we found in IHH cells cultured with insulin and PDGF-AA that an impaired INSR content is closely linked to reduced INSR tyrosine phosphorylation at residue Y972 (**Fig. 4d**). Inversely the PKCε mediated-threonine phosphorylation 1376 which inactivates the INSR, was increased in response to insulin or PDGF-AA (**Fig. 4e**).

To gain further insights into the intracellular mechanism through which PDGF-AA alters insulin signaling in hepatocytes, we performed RNA sequencing of IHH cells treated or not with insulin for 24 h. We found a profound dysregulation of expression of genes involved in both carbohydrate metabolism, inflammatory and insulin signaling pathways in response to insulin. Indeed, when we grew a network based on *PDGFA* through Ingenuity Pathway Analysis (IPA), we found a significant increase in the expression of genes of the VEGF and PDGF families, including as expected *PDGFA* (log2 Fold Change = 0.80; $p = 1.1 \times 10^{-11}$) (**Fig. S5 and Table S4**). Subsequently, we analyzed the diseases and/or functions highlighted by the insulin-evoked deregulated expressed genes in IHH cells. Among the significant outputs, we found a network related to the metabolism of carbohydrates that includes *PDGFA* ($p = 1.2 \times 10^{-6}$; **Fig. S5, Table S4**). We also identified in IHH cells cultured with insulin for 24 h a decrease in the expression of the insulin receptor substrate 1 (*IRS1*) gene (**Fig. 5a**, **Fig. S5 and Table S5**). The decreased *IRS1* expression by insulin was confirmed by western blotting (**Fig. 5b**) and mimicked by PDGF-AA (**Fig. 5c**). Silencing of *IRS1* expression using small interfering RNAs confirmed that the IRS1 abundance is critical for AKT activation in IHH cells in response to insulin (**Fig. 5d**). Defective IRS1 level can therefore account for the impaired insulin signaling caused by PDGF-AA in insulin resistant hepatocytes. The decrease of *IRS1* is mediated by PKC activity since its inhibition by the PKCϴ and PKCε, inhibitor sotrastaurin (39) prevented the reduction of IRS1 caused by hyperinsulinemia (**Fig. 5e**), and in contrast, PKC activation by (PMA) mimicked the effect of insulin (**Fig. 5f**).

Altogether, our data suggest a role for hepatocyte PDGF-AA in promoting further liver insulin resistance via the decrease of INSR and IRS1, and the activation of both PKCϴ and PKCε.

**Autocrine regulation of PDGF-AA on its expression and the beneficial effects of metformin**

PDGF-AA stimulates its own expression in the liver (40). It suggests that in insulin resistant hepatocytes PDGF-AA amplifies its secretion, thereby perpetuating insulin resistance. We found that culture of IHH cells with PDGF-AA stimulated *PDGFA* expression (**Fig. 6a**) and PDGF-AA secretion (**Fig. S6**). The effect of PDGF-AA on its expression was mediated by PDGFR as the PDGFR tyrosine kinase inhibitor ki11502 prevented the rise of *PDGFA* mRNA of cells exposed to either insulin or to human PDGF-AA (**Fig. 6b and 6c**). The overexpression of *PDGFA* by PDGF-AA may require PKC activation since the PMA mimicked both insulin and PDGF-AA effects on the *PDGFA* mRNA (**Fig. 6d**) and inversely, the inhibitor sotrastaurin, alleviated the rise of *PDGFA* (**Fig. 6e**). Altogether, it is likely that in T2D, hyperinsulinemia-induced PDGF-AA aggravates its overexpression and its consequences (**Fig. 6f**). The most prescribed T2D drug metformin inhibits PKCε (41) and has been specifically proposed for diabetic patients with NAFLD and hepatocarcinoma (HCC) (42). Furthermore, metformin reduces HCC incidence in diabetic patients in a dose-dependent manner (43). We found that metformin also efficiently abolished the expression of insulin-induced *PDGFA* mRNA (**Fig. 7a**), protein content (**Fig. 7b**) and secretion (**Fig. 7c**). Thus, a part of the effects of metformin on insulin sensitivity may be mediated by the reduction of *PDGFA* overexpression in hepatocytes.

**Discussion**

GWAS have only identified so far few genes involved in NAFLD (44) and the contribution of epigenetics to T2D liver dysfunction is still elusive. While we initially identified 381 differentially methylated regions (DMRs) in the liver from obese patients with T2D compared to normal glucose obese patients, we only observed one genome-wide significant DMR (cg14496282 within *PDGFA*), associated with the increase of *PDGFA* expression in cis, in obese patients with T2D after adjusting for liver steatosis and NASH. It suggests that the other DMRs are a consequence rather than a cause of the early stages of NAFLD. On the contrary, cg14496282 might be instrumental in the functional impairment of the liver in obesity associated T2D. Notably, we found that liver *PDGFA* cg14496282 hypomethylation and concomitant rise in liver *PDGFA* expression were also associated with systemic insulin resistance in non-diabetic obese patients but not with their glucose values. Elevated *PDGFA* expression was also reported in biliary atresia (45), and is a diagnostic and prognostic biomarker of cholangiocarcinoma that is a liver cancer associated with severe insulin resistance (but paradoxically not with obesity) (32, 46). Thus, PDGFA seems to be a liver marker of insulin resistance and of chronic hyperinsulinemia. Furthermore, the genetic data from our analysis of GRS related to insulin resistance, suggest a causative effect of plasma insulin levels on methylation level, hepatic expression and secretion of this growth factor. Therefore, the elevated *PDGFA* expression in human liver from obese subjects can be directly due to their severe hyperinsulinemia.

*PDGFA* encodes a dimer disulfide-linked polypeptide (PDGF-AA) that plays a crucial role in organogenesis (20). The overactivation of the PDGF-AA receptor signaling is involved in cirrhotic liver regeneration (21) and the chronic elevation of PDGF-AA in mice liver induces fibrosis (22). High PDGF-AA levels contribute to hepatic fibrogenesis by activating hepatic stellate cells in mice (23–25). In human, the association of increased *PDGFA* expression with liver steatosis and fibrosis observed in our study and in others supports a similar fibrogenic role in human liver (18,19), in which chronic hyperinsulinemia might be instrumental (47). We also believe that PDGF-AA contributes to the inhibitory effect of chronic hyperinsulinemia on hepatocyte insulin signaling via a feedback autocrine loop (**Fig. 6f**). We indeed showed that PDGF-AA stimulates its own induction via the activation of PKC. This vicious cycle perpetuates high PDGF-AA level and thereby worsens liver insulin resistance. Our data further suggests that the negative effect of PDGF-AA on insulin signaling is the consequence of the decrease of IRS1 and INSR, and PKC activation including PKCΘ and PKCε. These two kinases are known to phosphorylate IRS-1 and the insulin receptor on serine residues, that impairs the association of the insulin receptor with IRS proteins, leading to the blockade of AKT activation and of the downstream signaling (35,36).

Our findings may have a major interest for the treatment of T2D and of its hepatic complications. We showed that metformin, the most-widely prescribed oral insulin sensitizer agent prevented the PDGF-AA insulin-induced vicious circle. Metformin has been specifically proposed for diabetic patients with NAFLD and hepatocarcinoma (HCC) (42). Metformin reduces the risk of HCC incidence in diabetic patients in a dose-dependent manner (43). Thereby metformin may improve liver insulin sensitivity at least in part through PDGF-AA liver blockade, explaining its long-term effect against HCC. Beside liver insulin sensitizers, blocking PDGF-AA activity may be a promising alternative anti-diabetic therapeutic. The anti-tumor PDGFR inhibitor imatinib demonstrated unexpected (and unexplained until now) improvement of insulin sensitivity in insulin-resistant rats (48) as well as a dramatic blood-glucose-lowering effect in diabetic subjects treated for leukemia (49,50).

Our study also suggests that human epigenome analysis, when directly performed in disease-affected tissues is an efficient tool to make progress in the pathogenesis of common diseases. Furthermore, it opens avenues in the identification of new drug targets to combat T2D, and complications linked to insulin resistance, including NAFLD and cancer.

**Author contributions**

PF, SC, LY and AA designed the study. LY, AA, MC, AB and PF drafted and wrote the manuscript. LY and MC performed statistical analyses. AB, OS and IR performed the bioinformatics analysis. SL, JM, JD, GL, LR, MK, MT, JB, GQ, EA, SS, PJ, RB, SGC, VP, AL and MDC performed the experiments. AA, SC, MC, RC, VR, SL, JM, LR, GL, AL, TD, PM, BS, AS, JA, CP, JH, AB, FP and PF revised the manuscript. All authors have read and approved the final version of the manuscript.

**Acknowledgements**

This study was supported by nonprofit organizations and public bodies for funding of scientific research conducted in France and within the European Union: "*Centre National de la Recherche Scientifique*", "*Université de Lille 2*", "*Institut Pasteur de Lille*", "*Société Francophone du Diabète*", "*Contrat de Plan Etat-Région*", "*Agence Nationale de la Recherche*", ANR-10-LABX-46, ANR EQUIPEX Ligan MP: ANR-10-EQPX-07-01, European Research Council GEPIDIAB - 294785. We are grateful to Ms Estelle Leborgne for helping in the illustrations of the manuscript.

**Legends of Figures**

**Fig. 1. a)** Quantile-quantile (qq-) plot showing the residual inflation of test statistics before and after genomic-control correction. **b**) Manhattan plot centered on *PDGFA* cg14496282 methylation site showing association signal within *PDGFA* bounds. Hepatic *Pdgfa* expression in **c**) 6 weeks old male B6 mice that were diet-induced obese (DIO) responder (Resp, black circle) and DIO-non-responder (nResp, white circle) and in **d**) in BXD mice fed on Chow diet (CD, green circle) or a HFD (black circle) fed for 21 weeks. *** indicates p value < 0.0001 by unpaired t test with Welch's correction.

**Fig. 2. a**) PDGF-AA secretion from IHH cells and primary human hepatocytes was measured by ELISA kit. **b**) Measurement of insulin-induced AKT phosphorylation in response to human insulin (NovoNordisk) for the indicated times. IHH cells were incubated in a culture medium containing 5 mM Glucose, 2 % FCS with or without 100 nM human insulin for the indicated times. AKT phosphorylation was stimulated by 200 nM insulin for one hour. Immunoblotting for phospho-AKT (P-AKT) was done using the anti phospho-AKT (Serine 473) antibodies. The Fig. shows the result of a representative experiment out of three. **c**) Effect of insulin on the glycogen production. Insulin-induced glycogen production was measured by ELISA in IHH cells that were pre-cultured with 100 nM insulin for 16 h and 24 h. **d**) Increase of *PDGFA* mRNA by insulin. IHH cells were cultured with 100 nM human insulin for 16 h and 24 h. The *PDGFA* mRNA level was quantified by qRT-PCR and normalized against *GUSB*. The expression levels from untreated cells were set to 100 %. Data are the mean ± SEM (**: $p < 0.001$). **e**) PDGF-AA abundance in IHH cells cultured with insulin. IHH cells were cultured with 100 nM human insulin for the indicated times. PDGF-AA content was quantified by Western Blotting experiments. The blot is one representative out of three independent

experiments. **f**) PDGF-AA secretion in response to insulin. IHH cells were cultured with insulin for the indicated times. The measurement of PDGF-AA from the supernatant was achieved by ELISA. **g**) Methylation levels at *PDGFA* cg14496282 in response to insulin. IHH cells were cultured in a culture medium containing 5 mM Glucose, 2 % FCS with or without 100 nM human insulin for 24 h. Methylation level at the cg14496282 was quantified by the Infinium HumanMethylation450 BeadChip. **h**) Effect of chronic insulin in Insulin-induced AKT activation in HepG2 cells. HepG2 cells were incubated in a culture medium with 100 nM human insulin or without (Ctl) for 24 h. AKT phosphorylation was stimulated by 200 nM insulin for one hour. Immunoblotting for phospho-AKT (P-AKT) was done using the anti phospho-AKT (Serine 473) antibodies. The Fig. shows the result of a representative experiment out of three. **i**) Increase of *PDGFA* mRNA by insulin in HepG2 cells. The *PDGFA* mRNA level was quantified by qRT-PCR in HepG2 and IHH cells that were cultured with insulin for 24 h. The *PDGFA* mRNA was normalized against *GUSB*. The expression levels from untreated cells were set to 100 %. Data are the mean ± SEM (**: $p < 0.001$).

**Fig. 3.** Effects of **a**) human PDGF-AA recombinant, **b**) PDGFA blocking antibodies or **c**) the PDGFR inhibitor ki11502 on insulin-induced AKT activation. Activation of AKT was monitored by western blotting using total proteins from IHH cells that were cultured with the human recombinant PDGF-AA at the indicated concentrations for 24 h, which subsequently were incubated with 200 nM insulin for stimulating AKT phosphorylation. For **a-c**, IHH cells were co-incubated in a culture medium containing 5 mM Glucose, 2 % FCS with or without 100 nM human insulin for 24 hours plus **a**) PDGF-AA at the indicated concentration, **b**) PDGFA antibodies (+; 0.75 μg or ++; 1.5 μg) or **c**) ki11502 at the indicated concentration. The Figures show the result of a representative experiment out of three. **d**) Effect of the PDGFR inhibitor ki11502 on the glycogen production. Glycogen was measured by ELISA in IHH cells

that were co-cultured with 5 μM ki11502 and insulin for the indicated times. Glycogen was monitored after stimulating cells in a KRP buffer without (Ctrl) or with insulin for 1 hr and 20 mM glucose. Glycogen was monitored by ELISA.

**Fig. 4**: **a**) Volcano plot showing differences in putative serine/threonine kinase activities between control and insulin-treated IHH cells for 24 hrs. Specific and positive kinase statistic (in red) show higher activity in IHH cultured with insulin compared with control samples. Effects of **b**) insulin and **c**) PDGF-AA on the phosphorylation of PKCϴ and PKCε. IHH cells were cultured with insulin for the indicated times or PDGF-AA (for 24 hrs). Phosphorylation of PKCϴ (Ser 676) and PKCε (Ser 729) were measured by western blotting and normalized against total PKCϴ and PKCε. **d**) Effect of insulin and PDGF-AA on the tyrosine phosphorylation (Y972) of INSR (p-IR) and threonine phosphorylation of INSR (T1375), INSR abundance (IR) and αTubulin. Western Blotting experiments were achieved from total proteins of IHH cells cultured either with 100 nM insulin of 100 ng/ml PDGF-AA for 24 hr. Phosphorylation of INSR was done by stimulating IHH cells with insulin for 1 h. The Figures show the result of a representative experiment out of three.

**Fig. 5:** Effect of insulin on **a**) *IRS1* mRNA level and **b**) protein. The *IRS1* mRNA level and IRS1 abundance was quantified by qRT-PCR and Western Blotting in IHH cells cultured with 100 nM insulin for 24 h. The *IRS1* mRNA was normalized against *RPLP0*. The expression levels from untreated cells were set to 100 %. Data are the mean ± SEM (**: $p < 0.001$). **c**) effect of PDGF-AA on the IRS1 content. IHH cells were cultured with PDGF-AA at the indicated concentrations for 24 h. The Figures show the result of a representative experiment out of three. **d**) Effect of siRNA against IRS1 on the insulin-induced AKT activation. Duplexes

of small interfering RNAs were transfected in IHH cells for 48h. Thereafter, AKT phosphorylation on the serine 473 was induced with insulin for 1 h. The Figures show the result of a representative experiment out of three. Effect of **e**) the PKC inhibitor sotraustorin PKC and **f**) phorbol 12-myristate 13-acetate (PMA) PKC activator on the expression of *IRS1* mRNA. *PDGFA* mRNA was quantified in IHH cells cultured with either sotrastaurin at the indicated concentration in the presence of 100 nM insulin for 24 h, or PMA for the indicated times. The *IRS1* mRNA was normalized against *RPLP0*. The expression levels from untreated cells were set to 100 %. Data are the mean ± SEM (***: $p < 0.0001$).

**Fig. 6:** Effect of Ki11502 PDGFR inhibitor on the PDGFA expression in response to **a**) PDGF-AA and **b**) insulin. IHH cells were cultured for 24 h with 100 ng/ml PDGF-AA or 100 nM insulin in the presence or absence of 5 μM ki11502 for 24 h. Effect of **c**) PKC activator phorbol 12-myristate 13-acetate (PMA) and **d-e**) the PKC inhibitor sotraustorin on the expression of *PDGFA* mRNA. The *PDGFA* mRNA was quantified by qRT-PCR in IHH cells cultured with either PMA for the indicated times, or 100 nM insulin or 100 ng/ml PDGF-AA in the presence of 1 μM sotrastaurin or at the indicated concentration for 24 hrs. The *PDGFA* mRNA was normalized against *RPLP0*. The expression levels from untreated cells were set to 100 %. Data are the mean ± SEM (*, p<0.05; ***: $p < 0.0001$). **f**) Schematic representation of the mechanism linking hyperinsulinemia to hepatic insulin resistance in T2D. Insulin promotes hypomethylation and the rise of *PDGFA* expression, leading to PDGF-AA secretion. In turn, PDGF-AA inhibits the insulin signaling, in a negative autocrine feedback loop, via a mechanism involving a decrease in the IRS1 and INSR abundance and PKC (PKCϴ and PKCε) activation.

**Fig. 7**: Effect of metformin on the **a**) PDGFA mRNA level and PDGF-AA **b**) intracellular abundance and **c**) PDGF-AA secretion induced by 100 nM insulin for the indicated times.

**Fig. 1**

**a**

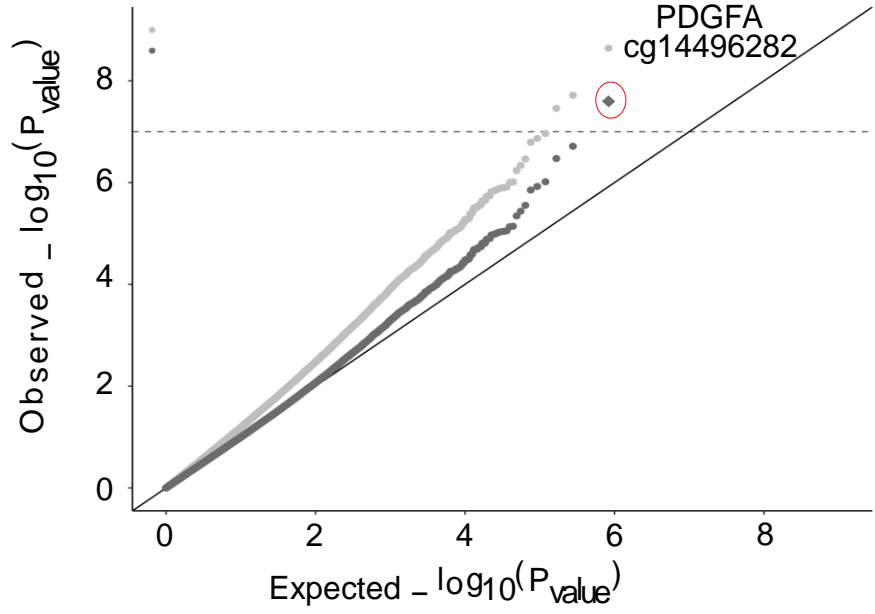

**b**

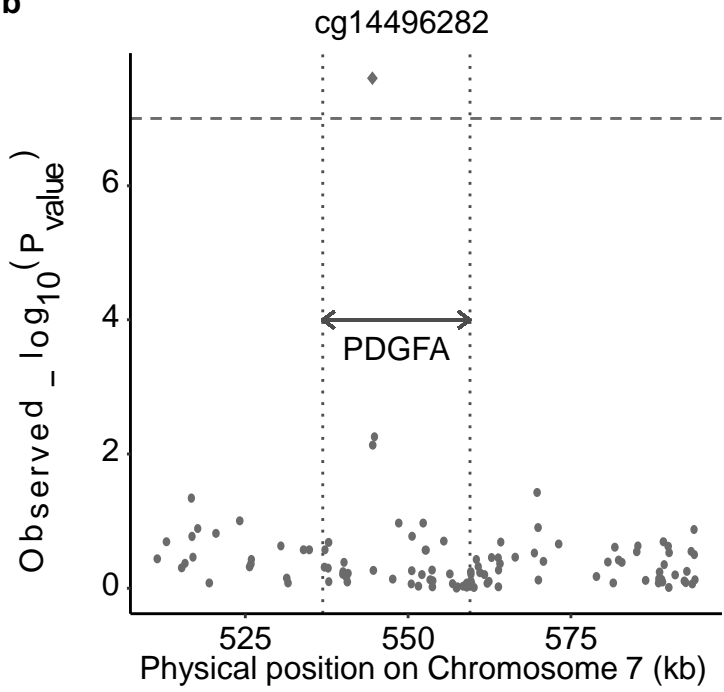

**Fig. 1**

**c**

Relative *Pdgfa* expression (Eef2 as endogenous control) plotted for nResp vs Resp. ***

**d**

*Pdgfa* (Liver mRNA) plotted for CD vs HFD. ****

**Fig. 2**

**a**

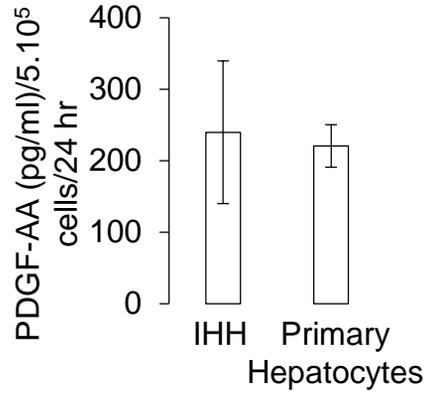

**b**

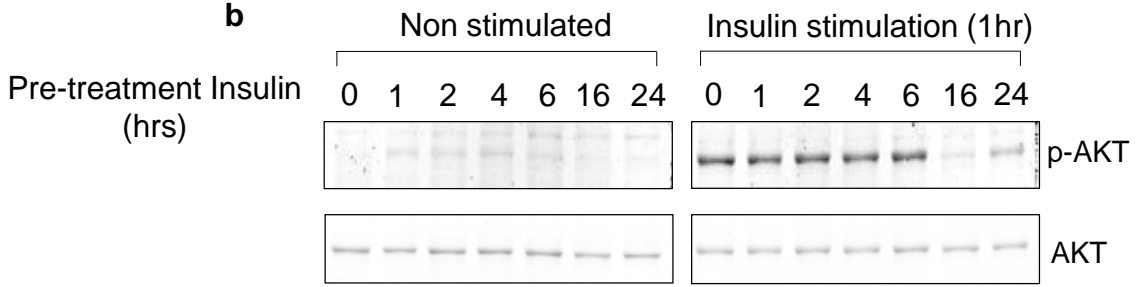

**c**

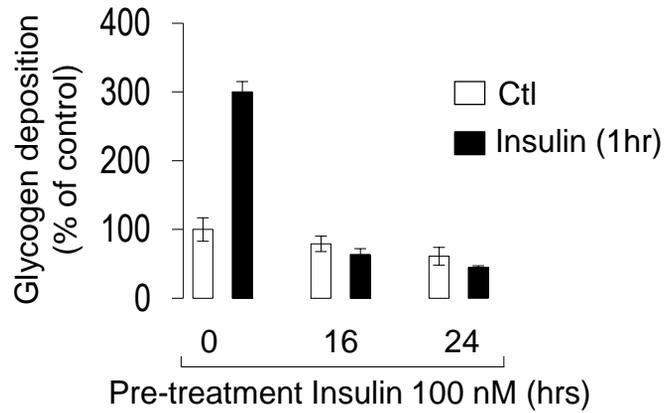

**Fig. 2**

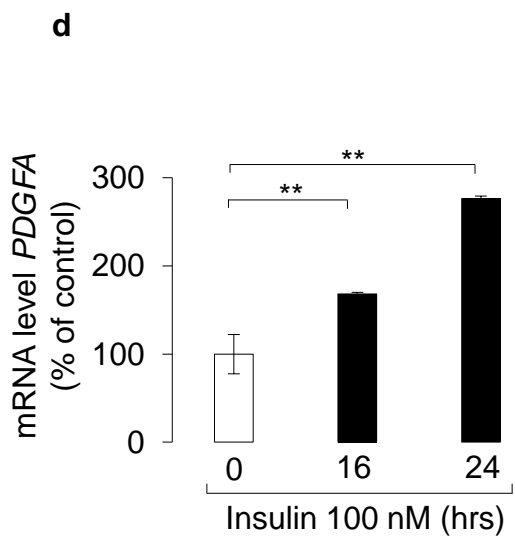

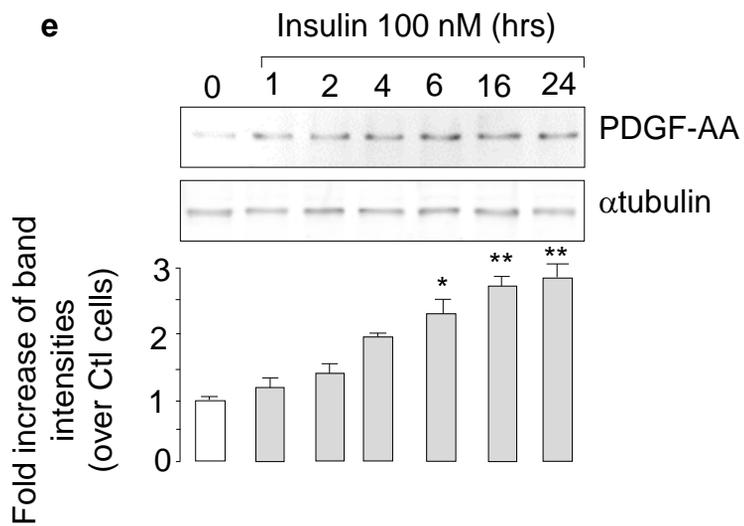

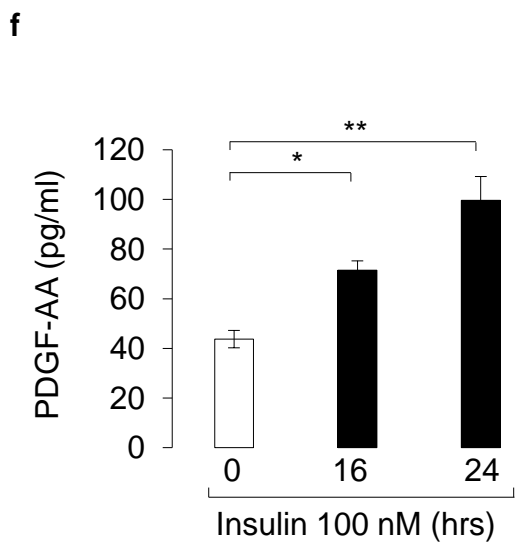

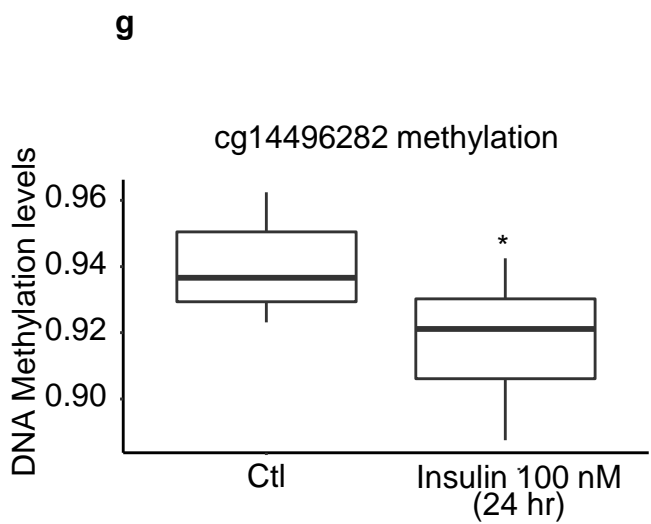

**Fig. 2**

**h**

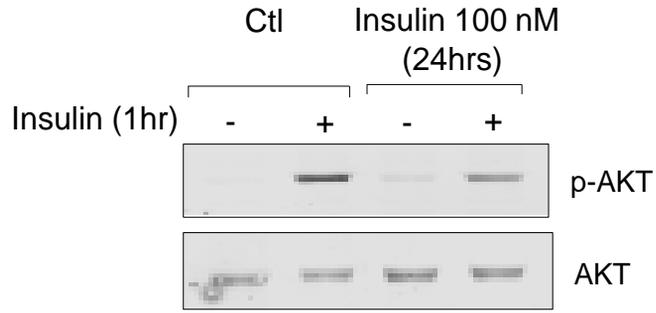

**i**

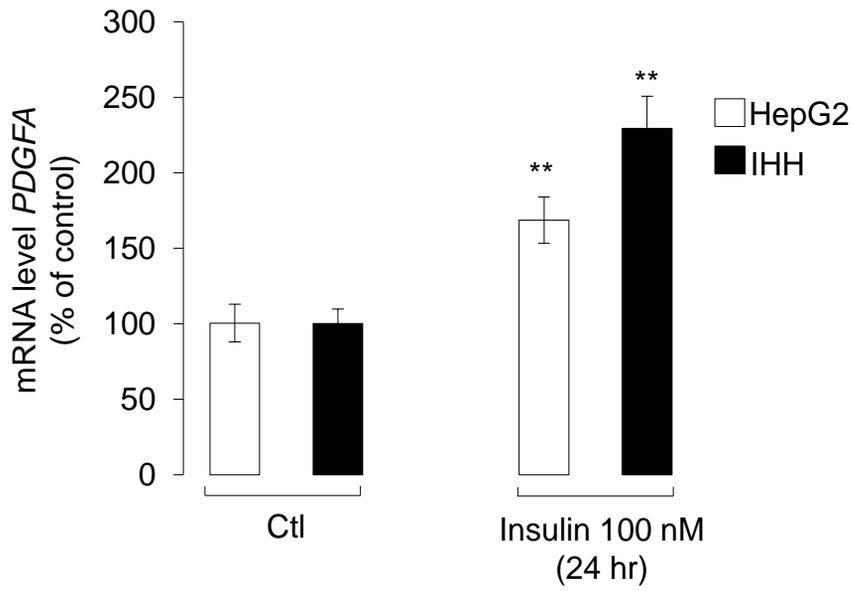

**Fig. 3**

**a**

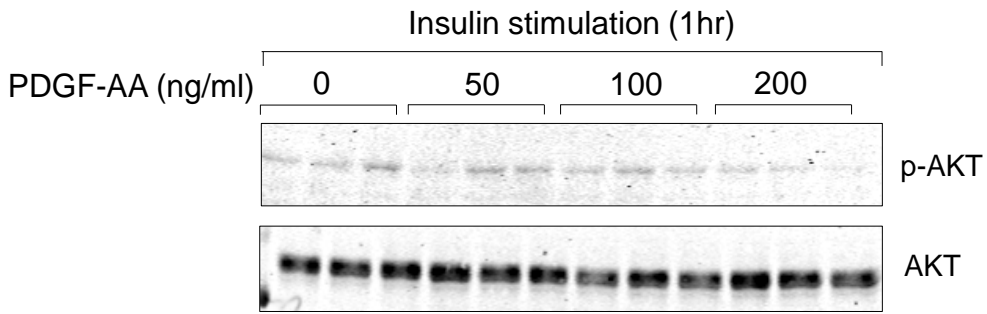

**b**

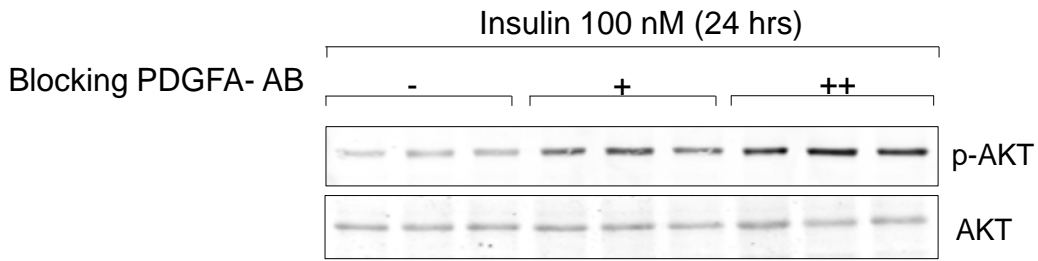

**Fig. 3**

**c**

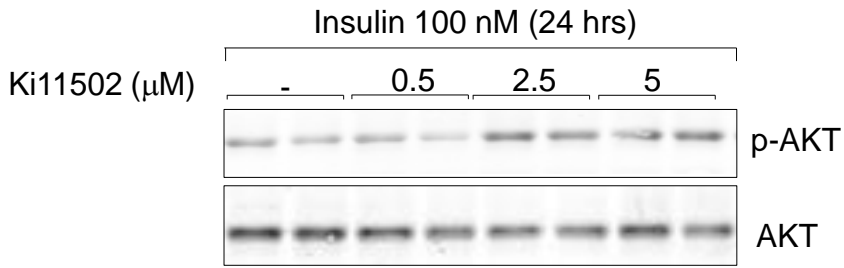

**d**

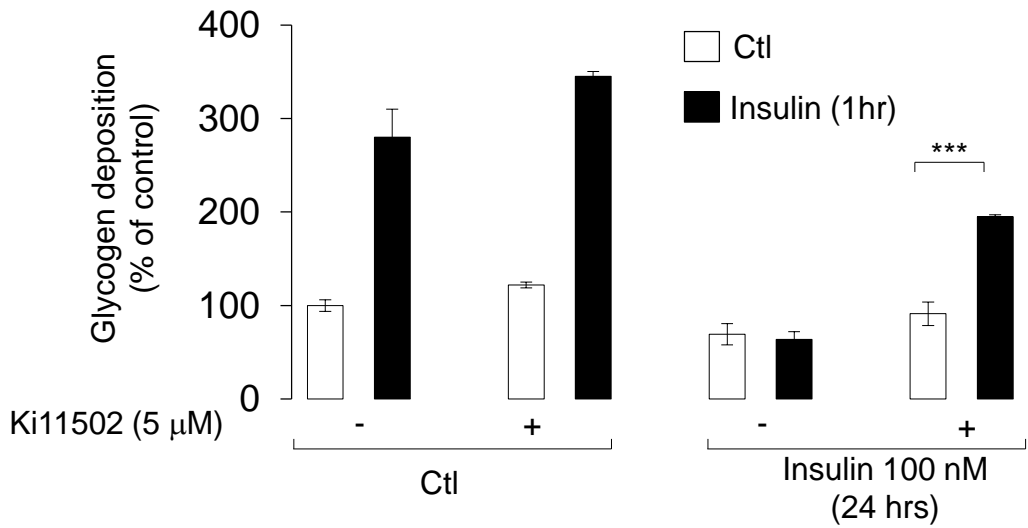

**Fig. 4**

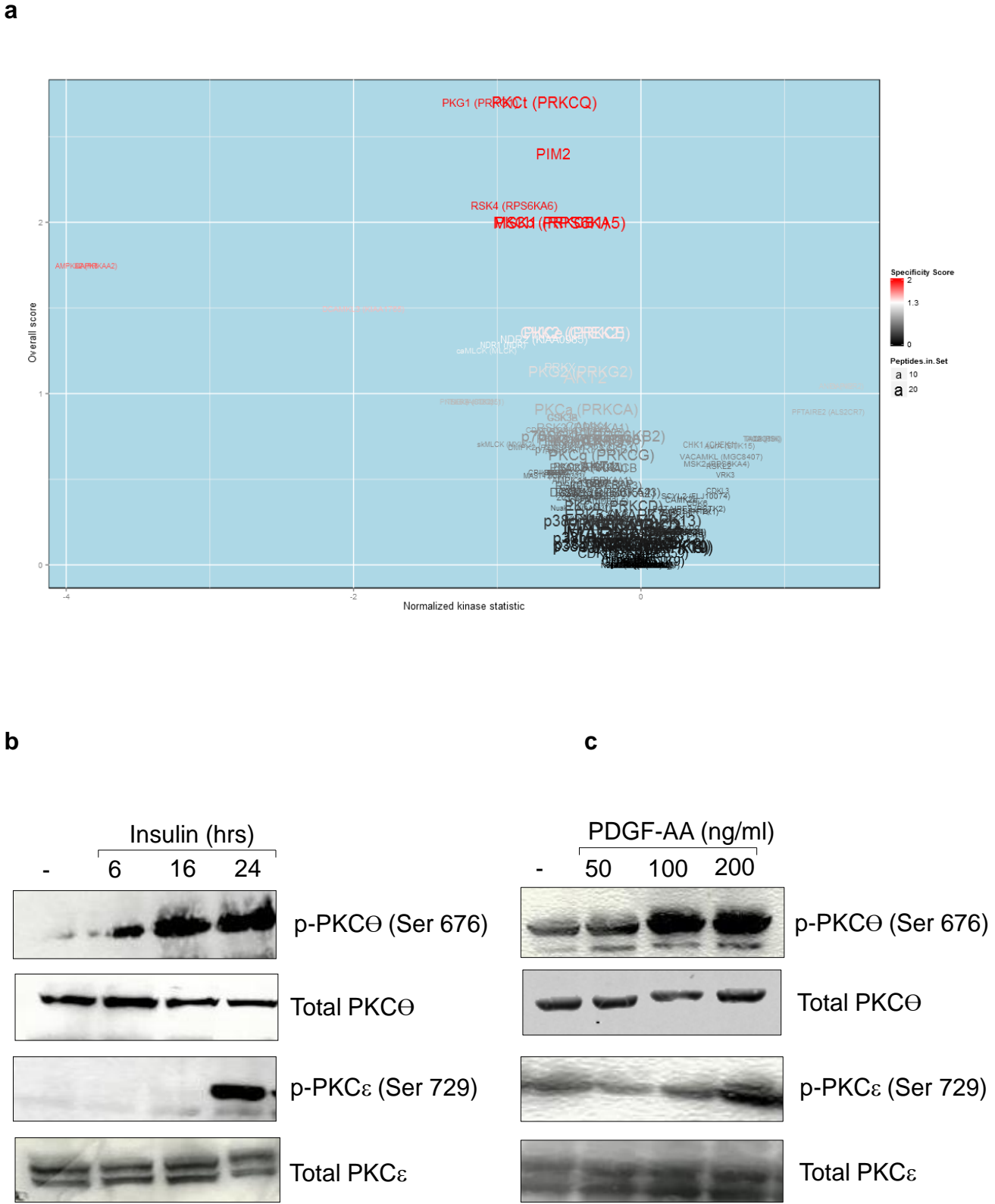

**Fig. 4d**

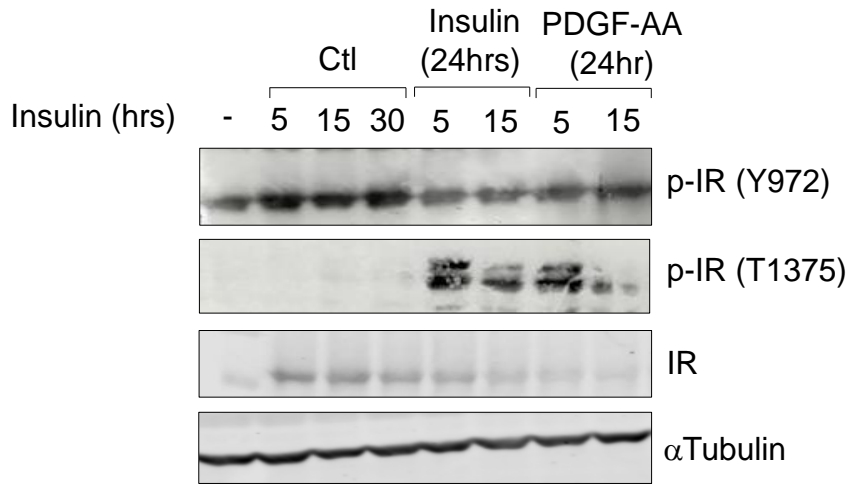

**Fig. 5**

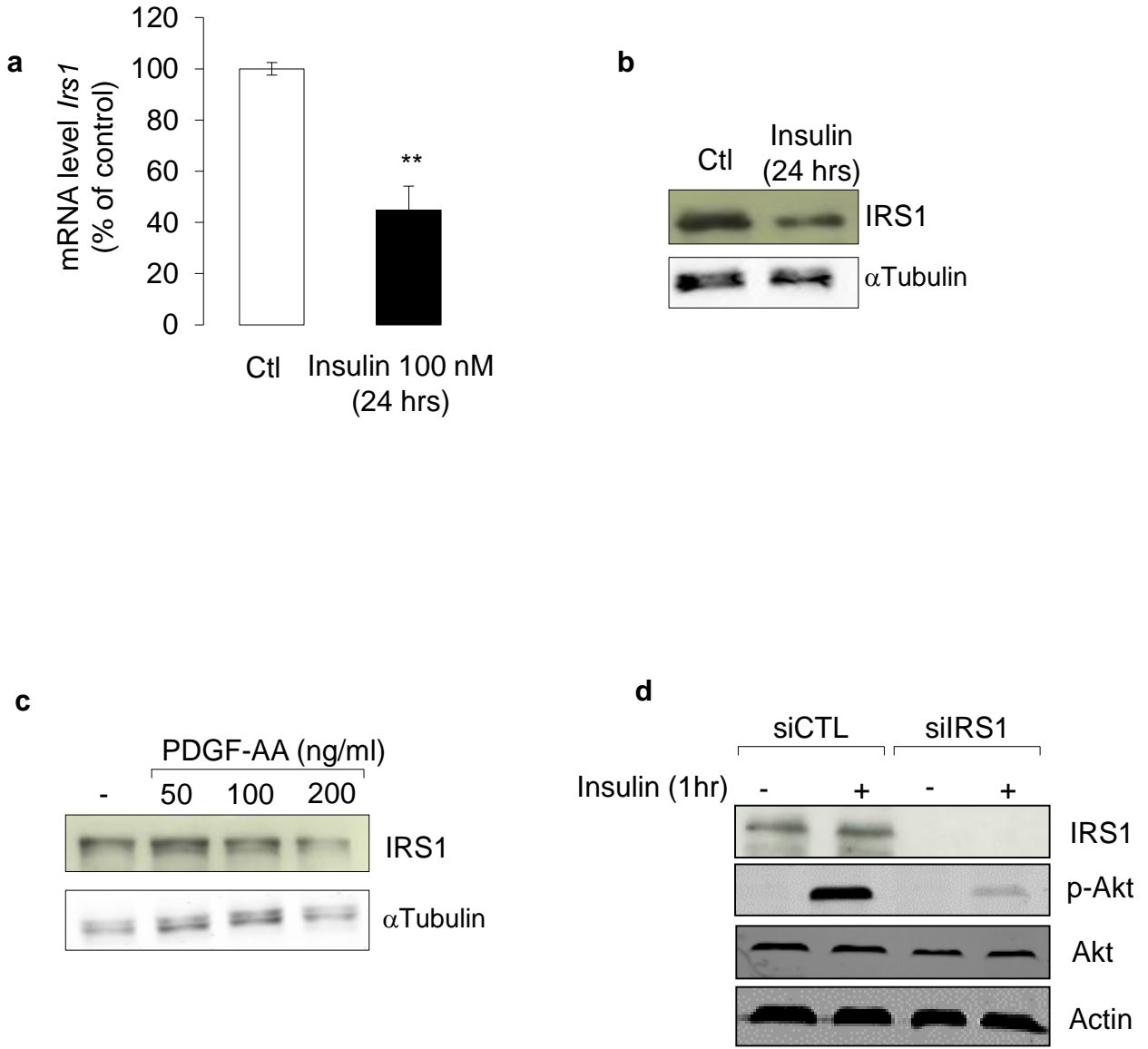

**Fig. 5**

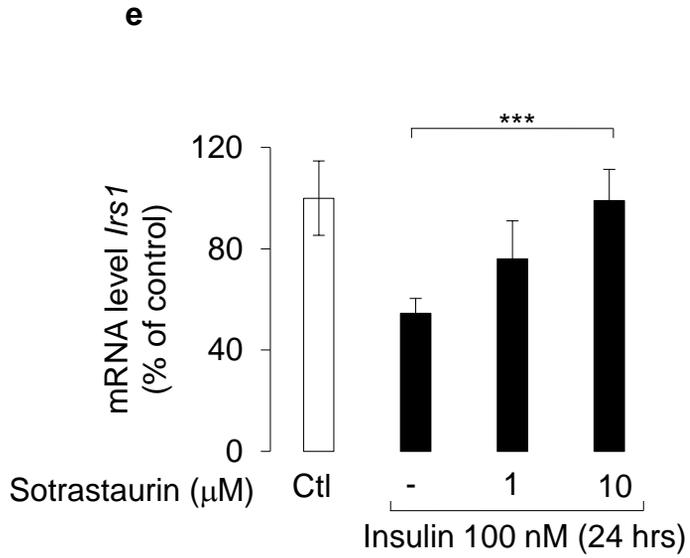

e

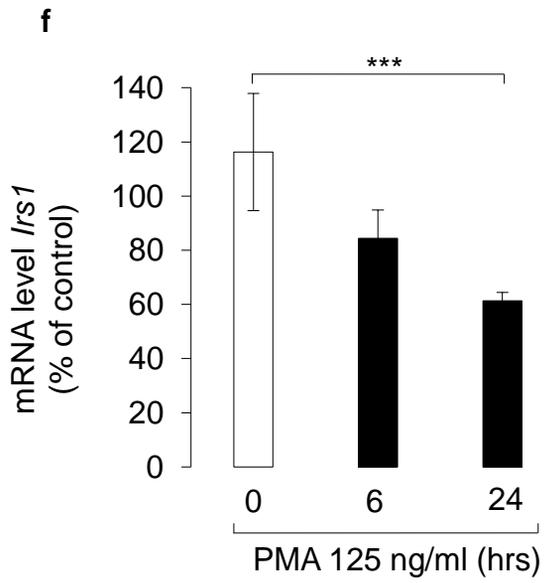

f

**Fig. 6**

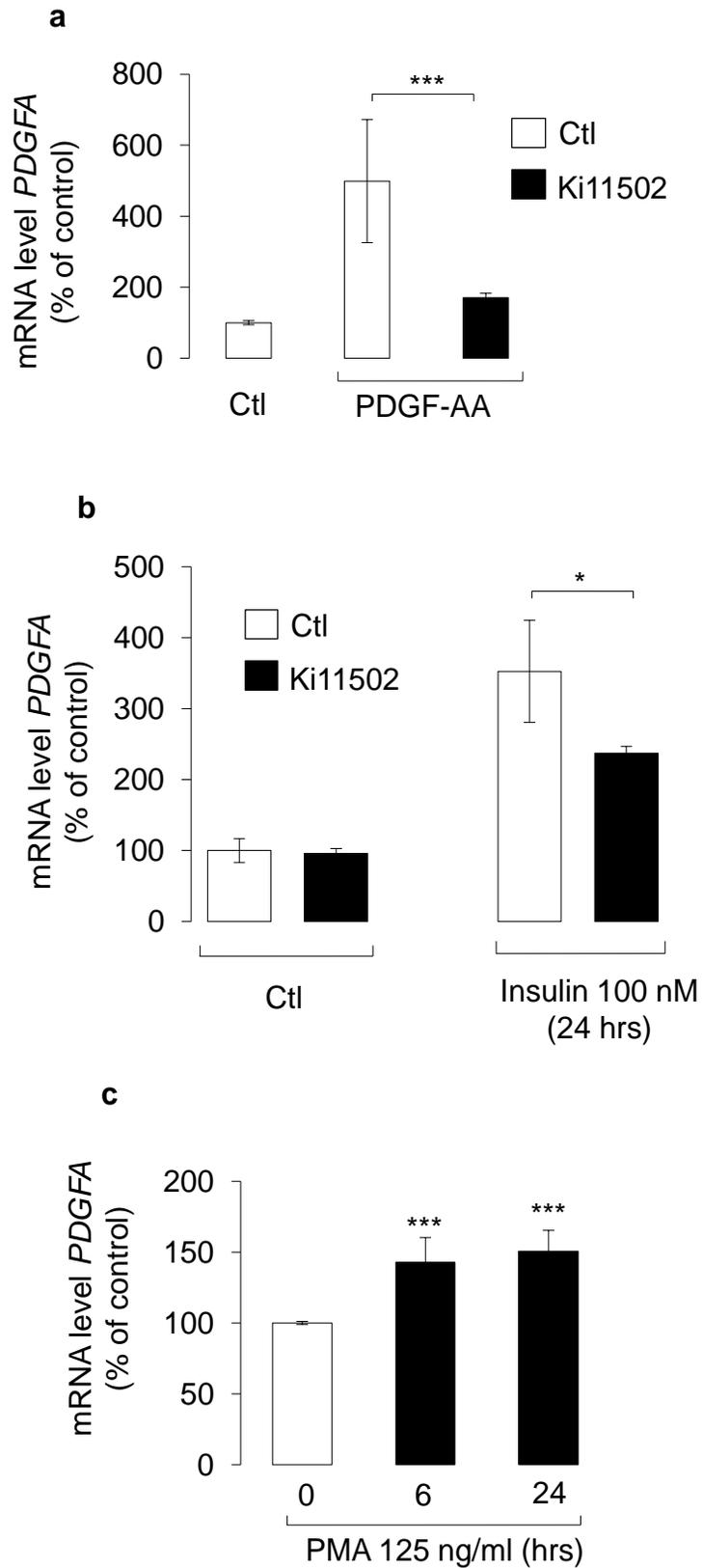

**Fig. 6**

**d**

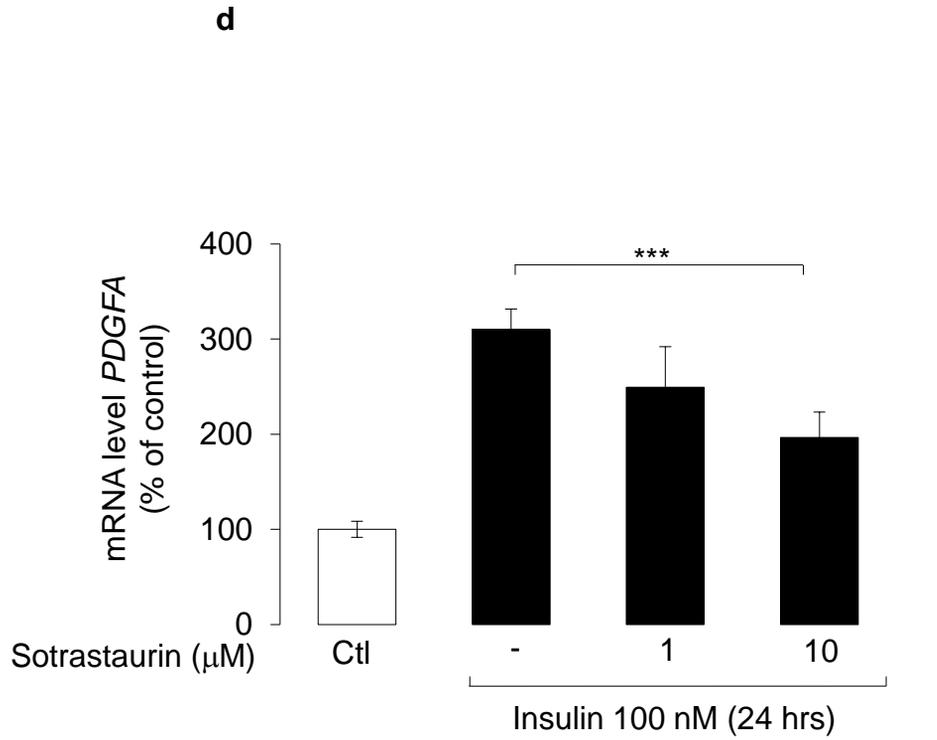

**e**

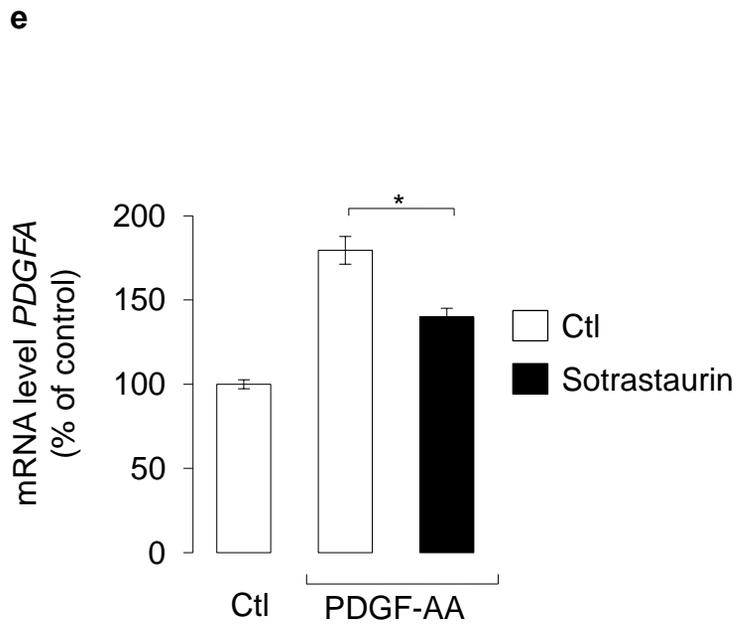

**Fig. 6f**

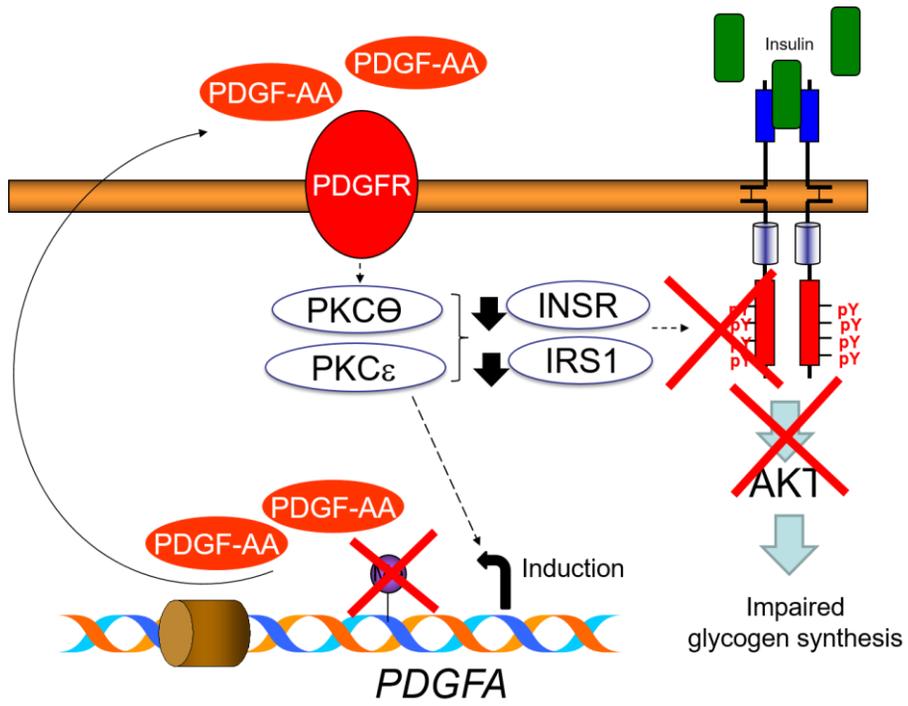

**Fig. 7**

a

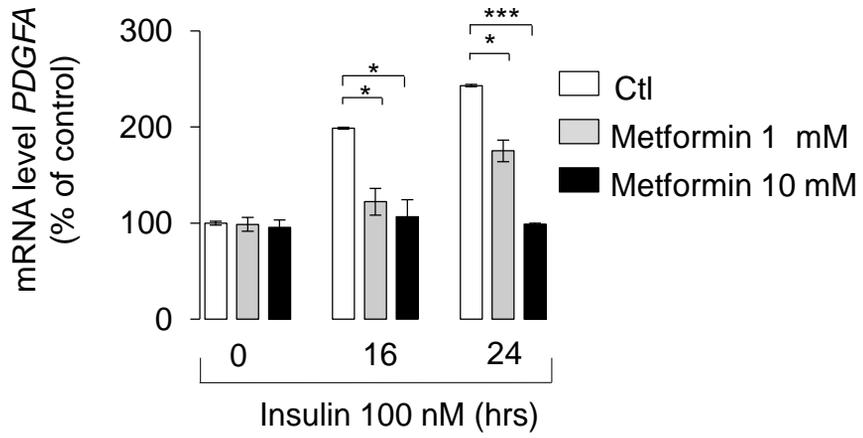

b

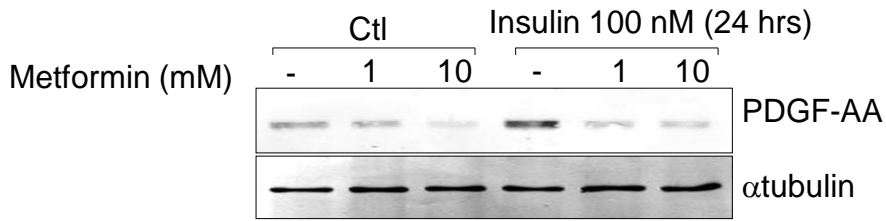

c

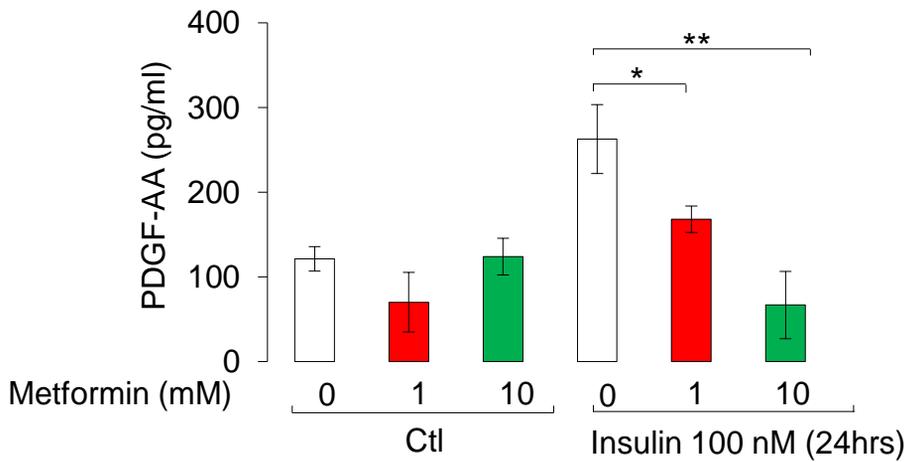

**Supplemental Information**

**Supplemental Experimental Procedures**

**Microarray mRNA expression analysis**. Transcriptome profiling was performed using the HumanHT-12 v4.0 Whole-Genome DASL HT Assay (Illumina). Total RNA was converted to cDNA using biotinylated oligo-dT18 and random nonamer primers, followed by immobilization to a streptavidin-coated solid support. The biotinylated cDNAs were then simultaneously annealed to a set of assay-specific oligonucleotides based on content derived from the National Center for Biotechnology Information (NCBI) Reference Sequence Database (release 98). The extension and ligation of the annealed oligonucleotides generate PCR templates that are then amplified using fluorescently-labeled (P1) and biotinylated (P2) universal primers. The labeled PCR products were captured on streptavidin paramagnetic beads, to yield single-stranded fluorescent molecules which were then hybridized, via gene-specific complementarity, to the HumanHT-12 BeadChip, thereafter fluorescence intensity was measured for each bead. Hybridized chips were scanned by using iScan (Illumina) and raw measurements were extracted by GenomeStudio software version 3.0 (Illumina).

**Epigenome-wide DNA methylation profiling**. Bisulfite converted DNA was amplified, fragmented and hybridized to the BeadChips following the standard Infinium protocol. All the samples were randomized across the chips and analyzed on the same machine by the same technician to reduce batch effects. After single base extension and staining, the BeadChips were imaged with the Illumina iScan. Raw fluorescence intensities of the scanned images were extracted with the GenomeStudio (V2011.1) Methylation module (1.9.0) (Illumina). The fluorescence intensity ratio was used to calculate the β-value which corresponds to the methylation score for each analyzed site according to the following equation: β-value = intensity of the Methylated allele (M) / (intensity of the Unmethylated allele (U) + intensity of the Methylated allele (M) + 100). DNA methylation β-values range from zero (completely unmethylated) to one (completely methylated). All samples had high bisulfite conversion efficiency (signal intensity >4000) and were included for further analysis based on GenomeStudio quality control steps where control probes for staining, hybridization, extension and specificity were examined. The intensity of both sample dependent and sample independent built in controls was checked for the red and green channels using GenomeStudio.

**Intermediate metabolic traits**. Body mass index (BMI) is the weight in kilograms divided by the *square* of the height in *meters*. The HOMA2-IR and HOMA2-B indexes were calculated using the HOMA (HOmeostasis Model Assessment) calculator based on fasting glucose and fasting insulin levels in subjects in steady-state situation (available at http://www.dtu.ox.ac.uk/homacalculator). We also calculated the Quantitative Insulin Sensitivity Check Index (QUICKI) using fasting glucose and fasting insulin levels (available at https://sasl.unibas.ch/11calculators-QUICKI.php). The degree of steatosis was defined as the percentage of hepatocytes containing fat droplets. In the control group of normoglycemic samples, the association levels reported in our study cannot be influenced by T2D treatment.

**SNP genotyping, ethnic characterization and genetic risk score.** For HapMap, 272 subjects including 87 of European ancestries [HapMap CEU], 97 of Asian ancestries [HapMap CHB] and 88 of African ancestries [HapMap YRI]) genotype calls at the 106,470 SNPs present on the Metabochip were available. The first two components were sufficient to discriminate ethnic origin (**Fig. S7**) and we observed that study participants clustered well with HapMap samples of European ancestries. We used 19 SNPs previously established for their association with fasting insulin to assess the possibility of causal direct link between DNA methylation at cg14496282. These 19 SNPs were on the Metabochip and all passed our quality control. These SNPs as well as the associated insulin raising alleles are reported in **Table S3**. We assessed the combined effect of these SNPs on DNA methylation using either fixed-effect meta-analysis and by testing the association of DNA methylation at cg14496282 with a genetic risk score (GRS) defined for each individual as the count of the number of fasting insulin raising alleles. We also tested the association between three other GRS (T2D, BMI and fasting glucose raising alleles) and DNA methylation at cg14496282 and expression in **Table S6**.

**DNA and RNA isolation.** The DNA samples were isolated using the Gentra Puregene Tissue Kit (Qiagen, Les Ulis, France). In order to improve isolation, DNA was then resuspended in Tris/Hcl/EDTA buffer and precipitated by adding *chloroform:isoamyl alcohol (4 %)*. Purity was determined by using a NanoDrop (NanoDrop Technologies, Wilmington, USA) and concentration was determined by using the Qubit® dsDNA BR Assay Kit (Life Technologies a brand of Thermo Fisher Scientific, Saint Aubin, France). The RNeasy® Lipid Tissue kit (Qiagen) and QIAamp RNA Blood Mini Kit (Qiagen) were used to isolate RNA from liver tissue and blood, respectively. Quality and integrity were tested on 2100 Bioanalyzer Instruments by using a RNA 6000 Nanochip (Agilent Technologies, Les Ulis, France). Concentration was determined by using the Qubit® RNA BR Assay Kit (Life Technologies).

**Oil Red O staining.** IHH cells were grown in a 24-well plate (at an initial density of $10^5$ cells/well) and incubated with 100 nM insulin for 16 hrs and 24 hrs. Cells were then washed three times with iced PBS and fixed with 4% paraformaldehyde for 30 minutes. After fixation, cells were washed three times and stained with Oil Red O solution for 15 min at room temperature. For Oil Red O content levels quantification, DMSO was added to sample; after shaking at room temperature for 5 min, the density of samples was read at 490 nm on a spectrophotometer

**Cell proliferation and apoptosis**. IHH cells were cultured with 100 nM insulin at different incubation times. Cell number was counted by trypan blue and apoptosis was determined by scoring cells displaying pycnotic or fragmented nuclei (visualized with Hoechst 33342). The counting was performed blind by two different experimenters.

**RNA-sequencing**. RNA samples with a RNA integrity number higher than 9.0 were extracted from IHH cells treated or not with 200 nM insulin for 24 hours, from at least three independent experiments. RNA libraries were prepared using the TruSeq Stranded mRNA Library Preparation Kit (Illumina, San Diego, CA, USA) following the manufacturer's instructions. The libraries were sequenced with the NextSeq 500 (Illumina). A mean of 50 million paired-end reads of 100 bp were generated for each sample. More than 92% of the reads for each library were effectively mapped to the hg19 human genome assembly using TopHat2. Subsequently, both quantification and annotation of the reads were performed using Rsubread. Finally, the differential gene expression analyses (*i.e.* 24 h insulin-treated IHH cells *versus* IHH cells at baseline) were performed using DESeq2. The differentially expressed genes were subjected to Ingenuity Pathway Analysis (Qiagen, Hilden, Germany) to decipher the major biological pathways and diseases emphasized by the significantly deregulated genes (with a *p*-value < 0.05).

**PamChip peptide microarrays for kinome analysis.** kinome analysis was achieved using Serine Threonine Kinases microarrays, which were purchased from PamGene International BV. Each array contained 140 target peptides as well as 4 control peptides. Sample incubation, detection, and analysis were performed in a PamStation 12 according to the manufacturer's instructions. Briefly, extracts from IHH cells cultured with insulin for 24 hours were made using M-PER mammalian extraction buffer (Thermo Scientific) for 20 minutes on ice. The lysates were then centrifuged at 15,871 $g$ for 20 minutes to remove all debris. The supernatant was aliquoted, snap-frozen in liquid nitrogen, and stored at –80°C until further processing. Prior to incubation with the kinase reaction mix, the arrays were blocked with 2% BSA in water for 30 cycles and washed 3 times with PK assay buffer. Kinase reactions were performed for 1 hour with 5 μg of total extract for the mouse experiment or 2.5 μg of total extract for the mature adipocyte and 400 μM ATP at 30°C. Phosphorylated peptides were detected with an anti-rabbit–FITC antibody that recognizes a pool of anti–phospho serine/threonine antibodies. The instrument contains a 12-bit CCD camera suitable for imaging of FITC-labeled arrays. The images obtained from the phosphorylated arrays were quantified using the BioNavigator software (PamGene International BV), and the list of peptides whose phosphorylation was significantly different between control and test conditions was uploaded to GeneGo for pathway analysis.

**Animal experiments.** The animal welfare committees of the DIfE as well as the local authorities (LUGV, Brandenburg, Germany) approved all animal experiments (reference number V3-2347-37-2011 and 2347-28-2014). All mice were housed in temperature controlled room (22±1 °C) on a 12:12 h light dark cycle (light on at 6:00 am) and had free access to food and water at any time. C57BL/6J breeding pairs (Charles River, Germany) as well as NZO/HIBomDife breeding mice (Dr. Kluge, German Institute of Human Nutrition, Germany) received a standard chow (SD; V153x R/M-H, Ssniff).

After weaning male C57BL/6J mice were fed a HFD (60 kcal% fat, D12492, Research Diets) ad libitum in groups of two to six animals per cage. Weekly body weight measurements were performed in the morning (8-10 am). Classification of mice either prone or resistant to diet-induced obesity was described before [Kammel et al. 2016 - PMID: 27126637]. At the age of 6 weeks, mice were killed after 6 h fasting. All tissues were directly frozen in liquid nitrogen and stored at -80 °C until further processing.

Female NZO mice received a standard chow after weaning and starting with 5 weeks of age, mice were switched to HFD containing 60 kcal% fat (D12492, Research Diets). Classification of diabetes-resistant (DR) mice in week 10 is based on assessment of liver density by computed tomography (LaTheta LCT-200, Hitachi-Aloka) combined with the measurement of blood glucose (CONTOUR® Glucometer, Bayer) [Lubura et al. 2015 - PMID: 26487005]. Liver density <55.2 Hounsfield units and blood glucose values of >8.8 and <16.6 mM both measured in week 10 predict later onset of diabetes with 83% and 84% probability, respectively. Combination of both parameters increased the prediction probability to 90 %.

Expression analysis. Total RNA from liver was isolated using QIAzol Lysis Reagent and RNeasy Mini Kit as recommended. Residual DNA was removed by DNase digestion using the RNase-Free DNase Set (QIAGEN). Subsequently, cDNA synthesis was conducted with 1 μg RNA, random hexamer primer as well as oligo(dT)$_{15}$ primer and M-MLV reverse transcriptase (Promega). qRT-PCR was performed with 12.5 ng cDNA in an Applied Biosystems 7500 Fast Real-time PCR system with PrimeTime® qPCR Probe Assay (IDT) for *Pdgfa* and the GoTaq® Probe qPCR Master Mix (Promega). Data were normalized to the expression of *Eef2* (TaqMan Gene Expression Assay, Life Technologies) as endogenous control. For genome expression profile either 4x44K or 8x60K whole mouse genome microarrays (Agilent Technologies) were used.

**Immunoblots analysis.** Proteins (40 μg) were subjected to SDS-PAGE analysis on 10% gels and transferred to nitrocellulose membranes. Rabbit polyclonal for total Akt, phospho-Akt (Ser-473 and Thr-308) were purchased from Cell Signaling. Monoclonal mouse β actin (clone AC-74; Sigma-Aldrich) was used as loading control.

**Primer sequences.** Primers for human *PDGFA* (sense 5′- GACCAGGACGGTCATTTACG -3′; antisense 5′- CGCACTCCAAATGCTCCT-3′). Primers for mouse *Pdgfa* (sense 5′- CAAGACCAGGACGGTCATTT -3′; antisense 5′- GATGGTCTGGGTTCAGGTTG-3′). Primers for human *RPLP0* (sense 5′- ACCTCCTTTTTCCAGGCTTT -3′; antisense 5′- CCCACTTTGTCTCCAGTCTTG -3′). Primers for human *IRS1* (sense 5′- ACAGGCTTGGGCACGAGT-3′; antisense 5′- AGACCCTCCTCTGGGTAGGA -3′)

**Glycogen measurement.** Total glycogen in IHH cells was determined by using Abcam Glycogen assay kit according to the manufacturer's instruction. The glycogen content was determined by using fluorometric assay and the fluorescence was measured using plate reader. Corrections for background glucose were made in all the samples and the corrected fluorescent readings were applied to standard curve and amount of glycogen in each well was determined. The glycogen content of each sample was normalized to their respective protein content and plotted using Graph Pad Prism 5

**Materials.** The PDGFRα inhibitor Ki11502, human PDGF-AA recombinant, Phorbol 12-Myristate 13-myristate (PMA) and Metformin were purchased from Sigma-Aldrich. The anti-PDGF-AA blocking antibodies were from Merckmillipore. The PKC inhibitor Sautrostaurin was from Selleckchem.

**Quantitative PCR.** Total RNA was extracted from IHH cells according to the manufacturer's protocol (RNeasy Lipid Tissue Kit, Qiagen). The RNA purity and concentration were determined by RNA Integrity Number (RNA 6000 Nano Kit, 2100 Bioanalyser, Agilent). Total RNA was transcribed into cDNA as described (17). Each cDNA sample was quantified by quantitative real-time polymerase chain reaction using the fluorescent TaqMan 5′-nuclease assays or a BioRad MyiQ Single-Color Real-Time PCR Detection System using the BioRad iQ SYBR Green Supermix, with 100 nM primers and 1 μl of template per 20 μl of PCR and an annealing temperature of 60 °C. Gene expression analysis was normalized against Beta-Glucuronidase (*GUSB*) expression or 60S acidic ribosomal protein P0 (*RPLP0*). The primer sequences are available in the supplemental information.

**Western Blotting and ELISA.** Cells were scrapped in cold PBS buffer and then cells pellet was incubated for 30 minutes on ice in the following lysis buffer (20 mM Tris acetate pH 7, 0.27 mM Sucrose, 1% Triton X-100, 1 mM EDTA, 1 mM EGTA, 1 mM DTT) supplemented with antiproteases and antiphosphatases (Roche, Meylan, France). Cell lysate was centrifuged 15 minutes at 18,000g and supernatant was collected as total proteins. For Western blotting experiments, 40 μg of total protein extract was separated on 10% SDS-Polyacrylamide gel and electrically blotted to nitrocellulose membrane. The proteins were detected after an overnight incubation of the membrane at 4°C with the specific primary antibodies against AKT (Santa Cruz Biotechnology, dilution 1:1000), PKCϴ (Abcam, dilution 1:1000), PKCε (Abcam, dilution 1:1000), IRS1 (Cell Signaling Biotechnology, dilution 1:1000), PDGF-AA (Merck Millipore, dilution 1:1000), αtubulin (Sigma, dilution 1:5000), phospho-AKT (Ser-473; Cell Signaling Biotechnology, dilution 1:1000), phospho-PKCϴ (Ser-676, Abcam, dilution 1:1000), phospho-PKCε (Ser-729, Abcam, dilution 1:1000), phospho-INSR (Tyr-972 and Thr-1375, Abcam, dilution 1:1000) in buffer containing 0.1% Tween 20 with either five percent BSA or five percent milk (for αTubulin). Proteins were visualized with IRDye800 or IRDye700 (Eurobio) as secondary antibodies. Quantification was performed using the Odyssey Infrared Imaging System (Eurobio). PDGFA released in the cell supernatant was quantified by ELISA kit (R &D Systems) according to the manufacturer's protocol.

**Table S1**. Clinical characteristics of the 192 samples (96 cases + 96 controls) included in the discovery cohort. Quantitative traits were compared between cases and controls using unadjusted linear regression and binary traits* were compared using Fisher exact test.

| Traits (unit) | Controls (n=96) Mean (SD) or n (%) | T2D cases (n=96) Mean (SD) or n (%) | p-value |
|---|---|---|---|
| cg14496282 methylation (%) | 60.3% (17.2%) | 41.3% (12.2%) | $9.27 \times 10^{-16}$ |
| PDGFA expression (SD) | 0.79 (0.35) | 1.04 (0.40) | $1.7 \times 10^{-6}$ |
| Age (years) | 46.7 (7.0) | 48.2 (6.3) | 0.13 |
| BMI (kg/m²) | 47.1 (7.4) | 49.1 (7.5) | 0.06 |
| Fasting glucose (mmol/l) | 5.2 (0.4) | 8.5 (3.2) | $2.2 \times 10^{-19}$ |
| Fasting insulin (pmol/l) | 78.7 (36.497) | 270.5 (854.2) | 0.029 |
| HOMA2-B (unitless) | 112.6 (34.8) | 71.5 (42.4) | $4.67 \times 10^{-11}$ |
| HOMA2-IR (unitless) | 1.5 (0.7) | 2.1 (1.435) | $8.0 \times 10^{-5}$ |
| QUICKI (unitless) | 0.34 (0.03) | 0.31 (0.04) | $3.83 \times 10^{-7}$ |
| Steatosis (%) | 21.3% (20.1%) | 42.4% (23.8%) | $4.77 \times 10^{-10}$ |
| NASH* (Yes/ No) | 4 (4.2%) | 17 (18%) | $4.44 \times 10^{-3}$ |
| Hepatic fibrosis* (Yes/ No) | 21 (22%) | 45 (47%) | $6.55 \times 10^{-4}$ |
| Alanine aminotransferase (UI/L) | 27.11 (15.8) | 34.3 (19.8) | $6.04 \times 10^{-3}$ |
| Aspartate aminotransferase (UI/L) | 22.344 (7.04) | 26.9 (14.3) | $5.5 \times 10^{-3}$ |

**Table S2**. Clinical characteristics of the 65 selected samples (12 cases + 53 controls) used for replication. Quantitative traits were compared between cases and controls using unadjusted linear regression and binary traits* were compared using Fisher exact test.

| Traits (unit) | Controls (n=53) Mean (SD) or n (%) | T2D cases (n=12) Mean (SD) or n (%) | p-value |
| --- | --- | --- | --- |
| cg14496282 methylation (%) | 63.3% (19.2%) | 41.4% (11.6%) | $3.59 \times 10^{-4}$ |
| Age (years) | 46.7 (13) | 54.6 (13) | 0.063 |
| Sex (female)* | 42 (79.2%) | 5 (71.4%) | 0.014 |
| BMI (kg/m²) | 40.5 (13.6) | 43.2 (11.5) | 0.542 |

**Table S3.** Association between fasting insulin raising alleles and DNA methylation at cg14496282. Associations were assessed using linear regression with methylation as response variable. All models were adjusted for age, BMI, total cholesterol, HDL cholesterol, triglycerides and fasting glucose. Effect sizes are reported as percentage of DNA methylation per allele.

| SNP | Closest gene (locus) | Fasting insulin raising allele | Effect size (p-value) in Controls | Effect size (p-value) in T2D cases | I² statistic (p-value) | Fixed effect meta-analysis effect size (p-value) |
|---|---|---|---|---|---|---|
| rs459193 | ANKRD55 | G | -1.85 (0.483) | -0.87 (0.691) | 0% (0.776) | -1.27% (0.448) |
| rs4865796 | MAP3K1 ARL15 | A | -3.53 (0.216) | 0.96 (0.656) | 37.49% (0.206) | -0.68% (0.692) |
| rs3822072 | FAM13A1 | A | -2.89 (0.267) | 0.12 (0.953) | 0% (0.353) | -0.97% (0.532) |
| rs1421085 | FTO | C | -0.66 (0.794) | -2.83 (0.124) | 0% (0.486) | -2.09% (0.158) |
| rs10195252 | GRB14 COBLL1 | T | -0.27 (0.913) | 0.22 (0.91) | 0% (0.876) | 0.03% (0.983) |
| rs1167800 | HIP1 | A | 2.31 (0.414) | 0.54 (0.793) | 0% (0.611) | 1.15% (0.486) |
| rs2943645 | IRS1 | T | -1.66 (0.552) | -0.73 (0.698) | 0% (0.783) | -1.02% (0.511) |
| rs4846565 | LYPLAL1 | G | -6.69 (0.006) | 1.49 (0.483) | 85.03% (0.01) | -2.15% (0.171) |
| rs6822892 | PDGFC | A | -0.88 (0.731) | -1.03 (0.62) | 0% (0.963) | -0.97% (0.546) |
| rs731839 | PEPD | G | -3.69 (0.197) | -0.95 (0.643) | 0% (0.432) | -1.88% (0.256) |
| rs17036328 | PPARG | T | -6.3 (0.122) | -4.42 (0.115) | 0% (0.701) | -5.02% (0.028) |
| rs2126259 | PPP1R3B | T | -0.24 (0.96) | -2.54 (0.418) | 0% (0.69) | -1.86% (0.477) |
| rs2745353 | RSPO3 | T | -4.72 (0.056) | -1.23 (0.546) | 17.56% (0.271) | -2.66% (0.088) |
| rs7903146 | TCF7L2 | C | -1.66 (0.619) | -2.75 (0.172) | 0% (0.78) | -2.46% (0.15) |
| rs974801 | TET2 | G | 4.32 (0.141) | -2.37 (0.229) | 72.58% (0.056) | -0.29% (0.861) |
| rs6912327 | UHRF1BP1 | T | -1.68 (0.576) | -3.87 (0.064) | 0% (0.546) | -3.16% (0.062) |
| rs1530559 | YSK4 | A | 3.94 (0.154) | -1.71 (0.356) | 65.92% (0.087) | 0.05% (0.974) |
| rs860598 | IGF1 | A | -5.39 (0.118) | 3.6 (0.095) | 79.95% (0.026) | 1.09% (0.548) |
| rs780094 | GCKR | C | 0.21 (0.936) | 1.35 (0.513) | 0% (0.736) | 0.92% (0.569) |
| Overall Meta-analysis (Cases + Controls) | - | - | -1.63% (0.011) | -0.83% (0.079) | 2.4% (0.428) | -1.11% (0.004) |
| Genetic Risk Score (GRS) | - | - | -1.68% (0.010) | -0.74% (0.099) | 30.2% (0.231) | -1.05% (0.004) |

**Table S4.** List of deregulated genes within the network of PDGFA

| Gene | Log2 Fold Change | p-value |
|---|---|---|
| PDGFRB | -0.68 | 3.2E-10 |
| IL1RN | -0.59 | 3.0E-15 |
| A2M | -0.33 | 2.2E-09 |
| FGF2 | -0.31 | 9.5E-03 |
| VDR | -0.25 | 3.2E-02 |
| PRKACB | -0.25 | 1.7E-03 |
| IRS1 | -0.21 | 1.1E-02 |
| POU5F1 | -0.19 | 6.0E-03 |
| NFATC2 | -0.17 | 2.7E-02 |
| NR3C1 | -0.14 | 4.4E-02 |
| PRKACA | -0.13 | 2.1E-02 |
| PDPK1 | -0.13 | 1.8E-02 |
| HDAC1 | -0.12 | 3.9E-02 |
| RXRB | -0.12 | 4.2E-02 |
| PRKAR1A | -0.12 | 2.4E-02 |
| SP1 | -0.11 | 3.9E-02 |
| NME2 | 0.11 | 2.6E-02 |
| SSRP1 | 0.12 | 2.2E-02 |
| SPP1 | 0.12 | 1.6E-02 |
| NCL | 0.12 | 8.2E-03 |
| TAF4 | 0.16 | 8.2E-03 |
| NFKB1 | 0.17 | 2.8E-03 |
| FURIN | 0.18 | 4.5E-02 |
| VEGFB | 0.20 | 1.5E-02 |
| PDAP1 | 0.21 | 3.0E-03 |
| HIF1A | 0.22 | 6.4E-05 |
| TGFB1 | 0.23 | 3.0E-04 |
| MAP2K1 | 0.23 | 1.3E-04 |
| MAP2K2 | 0.23 | 1.1E-02 |
| DNM2 | 0.25 | 1.7E-04 |
| RRAS | 0.25 | 4.4E-02 |
| TNFRSF1A | 0.25 | 7.6E-06 |
| GRB14 | 0.26 | 2.2E-02 |
| NFIC | 0.27 | 1.3E-02 |
| NME1 | 0.28 | 1.1E-06 |
| IL18 | 0.29 | 3.1E-02 |
| PDGFB | 0.43 | 1.2E-04 |
| PCSK5 | 0.52 | 8.3E-12 |
| EDN1 | 0.55 | 6.4E-05 |
| VEGFA | 0.56 | 5.8E-13 |
| NES | 0.64 | 1.2E-25 |
| DUSP1 | 0.67 | 9.7E-11 |
| PDGFA | 0.80 | 1.1E-11 |
| KLF5 | 0.87 | 3.6E-13 |
| EGR1 | 1.13 | 2.2E-16 |

**Table S5.** List of deregulated genes within the network related to the metabolism of carbohydrates

| Gene | Log2 Fold Change | p-value | Gene | Log2 Fold Change | p-value | Gene | Log2 Fold Change | p-value |
|---|---|---|---|---|---|---|---|---|
| G6PC | -1,69 | 2,3E-36 | FAS | -0,14 | 2,3E-02 | TCF7L2 | 0,24 | 5,7E-04 |
| PDGFRB | -0,68 | 3,2E-10 | CPT1A | -0,14 | 1,9E-02 | OAS1 | 0,25 | 2,3E-03 |
| IL1RN | -0,59 | 3,0E-15 | ALG2 | -0,14 | 1,2E-02 | SREBF1 | 0,25 | 8,4E-03 |
| GPAM | -0,54 | 1,5E-18 | ACADM | -0,14 | 2,2E-02 | TIGAR | 0,25 | 5,1E-04 |
| SLC2A2 | -0,53 | 3,8E-14 | PIGF | -0,14 | 3,5E-02 | HEXB | 0,25 | 1,1E-05 |
| CREB3L3 | -0,51 | 1,8E-05 | CHKB | -0,13 | 4,7E-02 | ITGB1 | 0,25 | 6,3E-06 |
| ENPP2 | -0,47 | 2,5E-04 | GM2A | -0,13 | 3,2E-02 | TNFRSF1A | 0,25 | 7,6E-06 |
| GNMT | -0,47 | 2,9E-04 | PDPK1 | -0,13 | 1,8E-02 | CTGF | 0,25 | 7,8E-03 |
| PPARGC1A | -0,46 | 1,3E-08 | EXTL2 | -0,12 | 4,2E-02 | CHST6 | 0,26 | 2,9E-02 |
| ETNK2 | -0,44 | 1,3E-09 | PHKB | -0,12 | 3,5E-02 | UGP2 | 0,26 | 3,0E-04 |
| SLC23A2 | -0,41 | 2,2E-14 | NAGA | -0,12 | 3,9E-02 | GCNT3 | 0,26 | 5,4E-04 |
| SH3YL1 | -0,39 | 6,4E-07 | MAN2A1 | -0,11 | 4,8E-02 | B3GNT3 | 0,27 | 8,0E-05 |
| PXYLP1 | -0,38 | 4,9E-04 | B3GAT1 | -0,11 | 3,5E-02 | GYS1 | 0,27 | 4,5E-05 |
| CYP3A5 | -0,35 | 6,1E-03 | PGM3 | -0,10 | 4,3E-02 | ADRBK1 | 0,28 | 2,0E-04 |
| TRPV1 | -0,35 | 2,0E-05 | NQO1 | 0,10 | 4,6E-02 | CSF1 | 0,29 | 3,5E-03 |
| PGAP1 | -0,34 | 1,7E-05 | RHOA | 0,10 | 3,8E-02 | IL18 | 0,29 | 3,1E-02 |
| PMM1 | -0,34 | 3,0E-06 | CLTC | 0,10 | 4,5E-02 | GPI | 0,29 | 2,4E-08 |
| NEU3 | -0,33 | 7,4E-05 | XBP1 | 0,10 | 2,9E-02 | NR1D1 | 0,30 | 3,1E-03 |
| PPP1R3F | -0,31 | 9,2E-03 | SCD | 0,11 | 3,6E-02 | GAL | 0,30 | 2,7E-02 |
| ST6GAL1 | -0,31 | 3,2E-10 | SPP1 | 0,12 | 1,6E-02 | B4GALNT1 | 0,30 | 5,2E-04 |
| FGF2 | -0,31 | 9,5E-03 | PTGFRN | 0,12 | 4,5E-02 | SFN | 0,31 | 1,5E-03 |
| GALNT7 | -0,29 | 3,6E-02 | GNB1 | 0,12 | 5,2E-03 | TPI1 | 0,31 | 5,5E-07 |
| RGN | -0,29 | 1,6E-05 | PTEN | 0,13 | 3,5E-02 | HBEGF | 0,31 | 2,1E-02 |
| PIGV | -0,27 | 9,9E-05 | SH3KBP1 | 0,13 | 2,9E-02 | IGF2 | 0,32 | 8,2E-05 |
| SLC37A4 | -0,27 | 1,2E-07 | DSE | 0,13 | 4,8E-02 | GOT1 | 0,33 | 4,0E-07 |
| PPARA | -0,27 | 4,3E-05 | GALE | 0,13 | 1,8E-02 | SLC2A8 | 0,33 | 2,3E-04 |
| GALT | -0,25 | 9,8E-05 | AFF4 | 0,13 | 4,2E-02 | PPARG | 0,33 | 2,7E-04 |
| IGF1R | -0,25 | 1,8E-05 | PRPS1 | 0,13 | 3,0E-02 | CHSY1 | 0,34 | 6,5E-08 |
| NDST2 | -0,25 | 1,2E-04 | CDC42 | 0,14 | 1,5E-02 | GMDS | 0,36 | 1,5E-09 |
| C5 | -0,25 | 2,0E-04 | SLC9A3R1 | 0,14 | 3,1E-02 | PPP1R3C | 0,37 | 1,8E-04 |
| GCKR | -0,25 | 1,0E-03 | TALDO1 | 0,14 | 7,0E-03 | NTSR1 | 0,37 | 4,9E-03 |
| ONECUT1 | -0,25 | 3,3E-02 | ARF6 | 0,15 | 7,0E-03 | GBE1 | 0,38 | 8,8E-06 |
| LIPC | -0,25 | 5,7E-03 | RPE | 0,15 | 1,7E-02 | DKK1 | 0,39 | 8,5E-08 |
| GPLD1 | -0,24 | 2,4E-03 | GNPDA1 | 0,15 | 1,3E-02 | SERINC2 | 0,39 | 3,3E-05 |
| ENPP1 | -0,24 | 3,9E-05 | GK | 0,15 | 4,2E-02 | FOXA2 | 0,39 | 2,8E-07 |
| NDST1 | -0,24 | 7,4E-04 | GLA | 0,15 | 1,3E-02 | PKM | 0,40 | 1,6E-12 |
| CTBS | -0,22 | 9,4E-03 | SH3GLB1 | 0,16 | 1,1E-02 | PLCH2 | 0,40 | 3,5E-03 |
| H6PD | -0,22 | 8,5E-04 | PYGB | 0,16 | 3,0E-02 | PGK1 | 0,42 | 5,4E-07 |
| IRS1 | -0,21 | 1,1E-02 | HS2ST1 | 0,17 | 4,7E-03 | PLAU | 0,43 | 9,5E-04 |
| ALDH5A1 | -0,21 | 2,8E-04 | GNE | 0,17 | 6,1E-03 | PDGFB | 0,43 | 1,2E-04 |
| S1PR2 | -0,21 | 3,0E-02 | CEBPB | 0,17 | 3,6E-02 | PLA2G3 | 0,44 | 5,4E-04 |
| KHK | -0,21 | 3,6E-03 | XYLT1 | 0,17 | 3,2E-02 | PPP1R3G | 0,45 | 1,6E-04 |
| HDAC5 | -0,20 | 3,4E-03 | COQ2 | 0,17 | 2,2E-02 | STBD1 | 0,45 | 2,6E-06 |
| PPP1R3E | -0,20 | 2,2E-03 | NFKB1 | 0,17 | 2,8E-03 | ALDOA | 0,45 | 6,8E-11 |
| SIAE | -0,20 | 7,8E-04 | FOXO1 | 0,18 | 2,4E-02 | ICAM1 | 0,45 | 1,2E-17 |
| PLCD1 | -0,20 | 4,9E-02 | G6PD | 0,18 | 3,0E-02 | ALDOC | 0,46 | 1,3E-10 |
| CHPT1 | -0,19 | 6,3E-04 | FUCA2 | 0,18 | 8,3E-04 | DUSP6 | 0,46 | 2,0E-15 |
| PIGZ | -0,19 | 2,9E-02 | TKT | 0,18 | 8,3E-03 | SLC2A1 | 0,49 | 1,3E-18 |
| SLC5A2 | -0,19 | 1,8E-02 | MTMR2 | 0,19 | 3,9E-03 | GRB10 | 0,49 | 1,8E-06 |
| CHST3 | -0,17 | 1,4E-03 | GYG1 | 0,19 | 5,5E-03 | PFKP | 0,52 | 3,8E-09 |
| EPHX1 | -0,17 | 3,3E-03 | INPP5E | 0,19 | 3,9E-02 | AGPAT2 | 0,53 | 3,1E-06 |
| FUCA1 | -0,17 | 1,3E-02 | ADORA2B | 0,20 | 1,2E-02 | SLC16A3 | 0,54 | 2,7E-10 |
| LCAT | -0,17 | 4,3E-02 | MYC | 0,20 | 2,3E-04 | EDN1 | 0,55 | 6,4E-05 |
| PIGP | -0,16 | 3,2E-02 | TRPV2 | 0,20 | 3,7E-03 | ANGPTL8 | 0,56 | 1,6E-06 |
| SULF2 | -0,16 | 3,1E-02 | NEU1 | 0,21 | 1,5E-05 | HK1 | 0,57 | 2,1E-05 |
| SERINC5 | -0,16 | 2,6E-02 | GAPDH | 0,21 | 1,5E-04 | CHST15 | 0,62 | 1,6E-10 |
| TGFBR1 | -0,16 | 1,3E-02 | PLCG2 | 0,21 | 2,7E-02 | CEMIP | 0,71 | 1,7E-20 |
| HECTD4 | -0,16 | 4,8E-02 | PRKAA2 | 0,21 | 2,2E-03 | PFKFB3 | 0,71 | 1,3E-07 |
| DYRK2 | -0,16 | 2,8E-02 | HS6ST1 | 0,22 | 7,7E-03 | IGFBP3 | 0,78 | 1,4E-08 |
| CPS1 | -0,15 | 1,1E-02 | HIF1A | 0,22 | 6,4E-05 | PPP1R15A | 0,80 | 3,5E-19 |
| PLCB1 | -0,15 | 2,7E-02 | PFKFB2 | 0,22 | 5,3E-04 | PDGFA | 0,80 | 1,1E-11 |
| EPM2AIP1 | -0,15 | 1,2E-02 | CHKA | 0,22 | 1,9E-05 | HK2 | 0,82 | 8,4E-23 |
| SRD5A3 | -0,14 | 3,9E-02 | TGFB1 | 0,23 | 3,0E-04 | NR4A1 | 0,86 | 1,0E-21 |
| PLSCR1 | -0,14 | 2,9E-02 | | | | | | |

**Table S6**. Correlation between multiple genetic risk scores (GRS) and PDGFA methylation in 192 ABOS study participants. GRS are calculated as the number of trait/risk increasing alleles over $N_{SNP}$ (number of SNPs) independent loci.

| Traits | Publication identifying genome-wide associated SNPs | Correlation between GRS and PDFGA DNA methylation at cg14496282 (p-value) |
|---|---|---|
| Type 2 diabetes ($N_{SNP}$=65) | Morris et al. Nat. Genet. (2012) | -0.07 (0.32) |
| BMI ($N_{SNP}$=97) | Locke et al. Nature (2015) | -0.01 (0.87) |
| Fasting glucose ($N_{SNP}$=24) | Vaxillaire et al. Diabetologia (2014) | -0.05 (0.51) |
| Fasting insulin ($N_{SNP}$=19) | Scott et al. Nat. Genet. (2012) | -0.21 (0.005) |

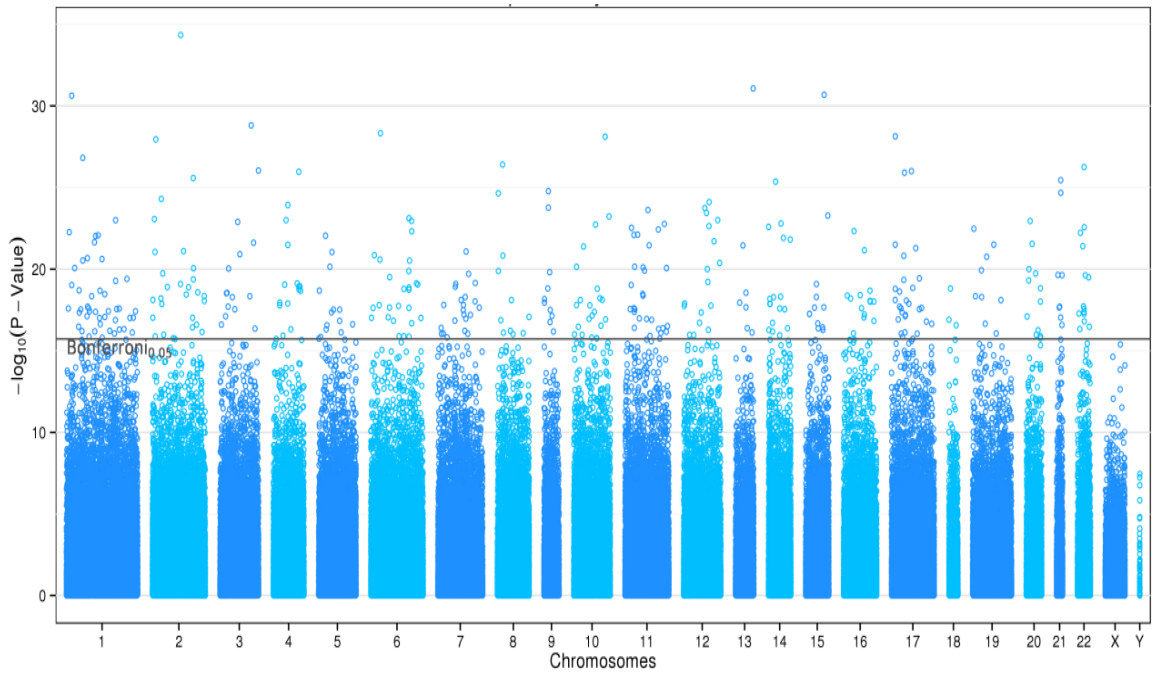

**Fig. S1.** CpG sites differentially methylated in livers of obese diabetic vs normoglycemic individuals. Liver methylome of 192 individuals revealed 381 differentially methylated CpG sites in 96 obese diabetic vs 96 normoglycemic.

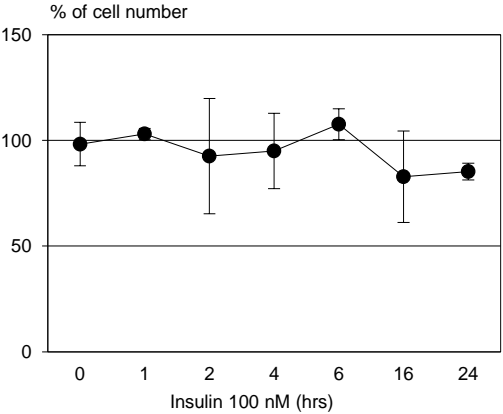 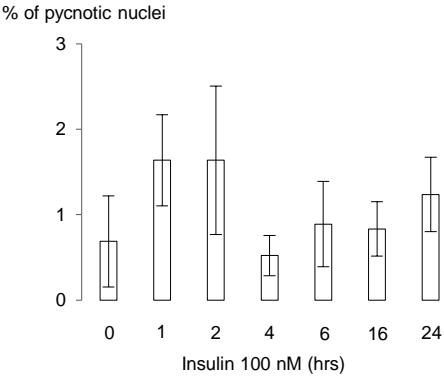

**Fig. S2. Cell proliferation and death in response to insulin**. IHH Cells were cultured with 100 nM insulin for the indicated times. **a**) Cell number and **b**) apoptosis were counted.

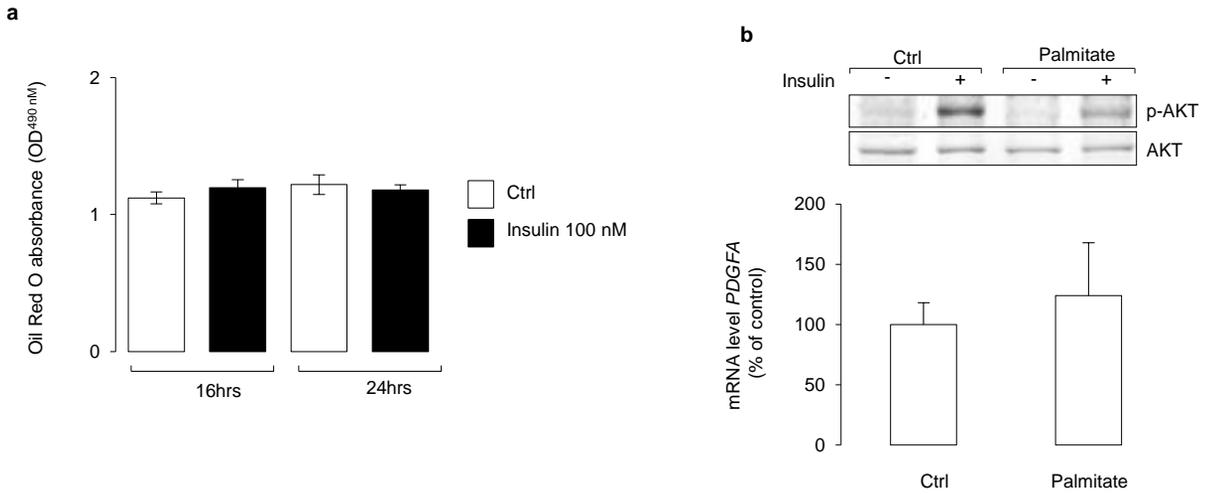

**Fig. S3. Role of lipids in IHH cells a) Lipid accumulation in IHH cells**. Cells were cultured with 100 nM insulin for the indicated times. Intracellular fat drops were read with spectrophotometer at 490 nm with Oil Red O staining, representing the mean ± SD of 3 independent experiments. **b) effect of palmitate on the *PDGFA* mRNA level.** IHH cells were cultured with 0.5 mM Palmitate (coupled to BSA) for 24 hrs. Impaired insulin-induced Akt activation was monitored by western blotting experiments (upper panel) using the anti-phospho-AKT (Thr-308) and anti-Akt antibodies. The The *PDGFA* mRNA level was quantified by qRT-PCR and normalized against *GUSB*. The expression levels from untreated cells were set to 100 %. Data are the mean ± SEM (n=3 independent experiments).

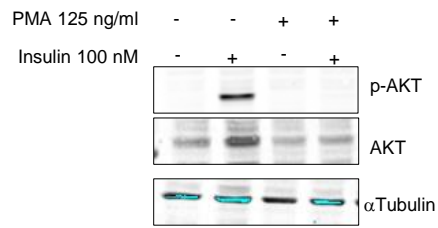

**Fig. S4. Effect of PMA on a) insulin-induced AKT activation**. IHH cells were incubated in a culture medium containing 5 mM Glucose, 2 % FCS with or without 100 nM human insulin and 125 ng/ml PMA for 30 minutes. Immunoblotting was done using the anti-phospho-AKT (Thr-308), anti-AKT and anti αTubulin antibodies.

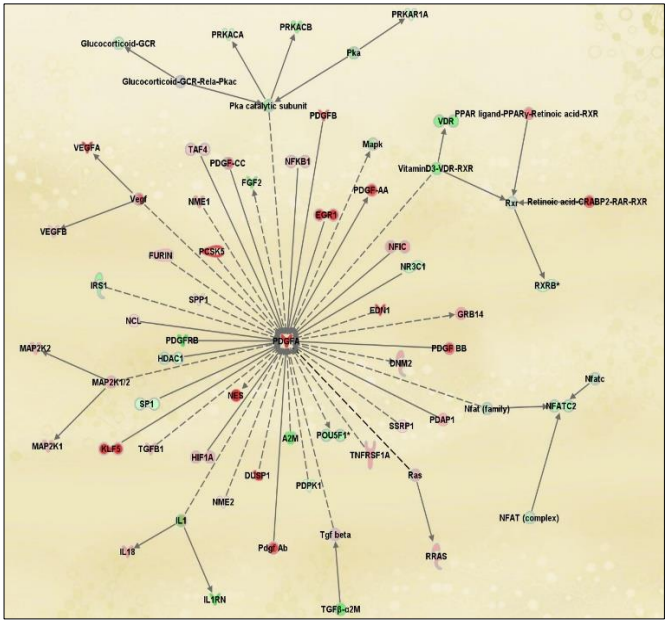 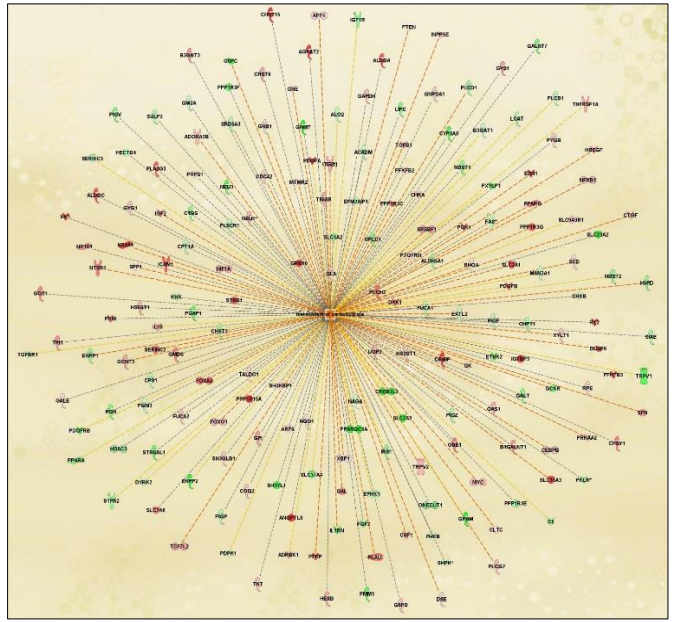

**Fig. S5. a) Dysregulated genes within the network of PDGFA.** Red and green colors mean increased and decreased gene expression, respectively. Solid lines embody direct interaction, while dotted lines embody indirect interactions. **b) Dysregulated genes within the network related to the carbohydrates metabolism.** Red and green colors mean increased and decreased gene expression, respectively. Each line embodies a predicted relationship: an orange line leads to activation, a blue line leads to inhibition, a yellow line highlights that the findings are inconsistent with state of downstream molecule, and a gray line highlight that the effect is not predicted.

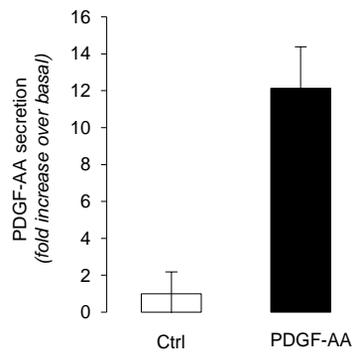

**Fig. S6. Effect of exogenous PDGF-AA on the PDGF-AA secretion**. IHH cells were cultured with 100 ng/ml human PDGF-AA or vehicle (Ctrl) for 24 hrs. PDGF-AA in the supernatant was measured by ELISA.

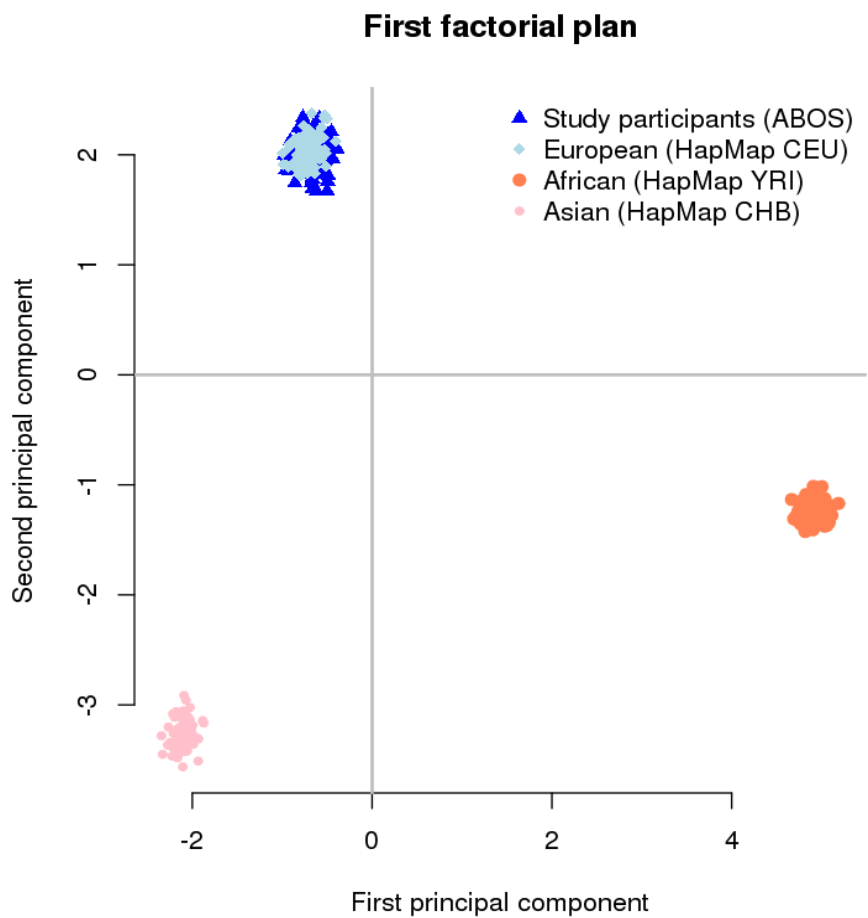

**Fig. S7. HapMap samples of European ancestries.** Principal component analysis on a combined dataset including 192 study participants and 272 samples from HapMap Project database and showing that study participants clustered well with HapMap samples of European ancestries (CEU).